\documentclass[10pt]{article}
\usepackage{graphicx}
\usepackage{amsmath}
\usepackage{amssymb}
\usepackage{caption2}
\begin{document}
\begin{center}
{\large {\bf \sc{  Analysis of the hidden-charm tetraquark mass spectrum with the QCD sum rules }}} \\[2mm]
Zhi-Gang  Wang \footnote{E-mail: zgwang@aliyun.com.  }     \\
 Department of Physics, North China Electric Power University, Baoding 071003, P. R. China
\end{center}

\begin{abstract}
In this article,  we take the pseudoscalar, scalar, axialvector, vector, tensor (anti)diquark operators as the basic constituents, and construct
  the scalar, axialvector and tensor tetraquark currents to study the  mass spectrum of the ground state hidden-charm tetraquark states  with
the QCD sum rules in a comprehensive way. We revisit the assignments of the $X$, $Y$, $Z$ states, such as the $X(3860)$, $X(3872)$, $X(3915)$,  $X(3940)$, $X(4160)$, $Z_c(3900)$, $Z_c(4020)$, $Z_c(4050)$, $Z_c(4055)$, $Z_c(4100)$, $Z_c(4200)$, $Z_c(4250)$, $Z_c(4430)$, $Z_c(4600)$, etc in the  scenario of tetraquark  states in a consistent way based on the QCD sum rules. Furthermore, we discuss the feasibility of applying the QCD sum rules to study the tetraquark states and tetraquark molecular states (more precisely, the color-singlet-color-singlet type tetraquark states), which  begin to receive contributions at the order   $\mathcal{O}(\alpha_s^0)$, not at the order $\mathcal{O}(\alpha_s^2)$.
 \end{abstract}



\section{Introduction}
In 2003, the  Belle collaboration   observed   a narrow charmonium-like state $X(3872)$ in the $\pi^+ \pi^- J/\psi$ mass spectrum in the exclusive decays  $B^\pm \to K^\pm \pi^+ \pi^- J/\psi$  \cite{X3872-2003}, which cannot be accommodated in the conventional two quark model as the $\chi_{c1}^\prime$ state with the quantum numbers  $J^{PC}=1^{++}$. Thereafter, about twenty   charmonium-like states were observed  by the  BaBar, Belle, BESIII, CDF, CMS, D0, LHCb collaborations \cite{PDG},   which cannot be accommodated in the conventional two quark model, and are denoted as the $X$, $Y$ and $Z$ states now, some are still needed confirmation and the quantum numbers have not been established yet. In Table \ref{Exp-XYZ-Table}, we  list out the masses, widths and $J^{PC}$ of the $X$, $Y$, $Z$ states in the $\bar{c}c$ region from the Particle Data Group \cite{PDG}. In 2018, the LHCb collaboration observed an evidence for an exotic   $\eta _c {{\pi } ^-}$  resonant state (now it referred to as $Z_c(4100)$) in the ${{B} ^0} \!\rightarrow \eta _c {{K} ^+} {{\pi } ^-}$ decays with   the significance of  more than three standard deviations,  the possible spin-parity assignments are $J^P =0^+$ and $1^-$ \cite{LHCb-Z4100}. We add the $Z_c(4100)$ in Table 1.

\begin{table}
\begin{center}
\begin{tabular}{|cccccc|} \hline\hline

    State    &M~(MeV)                            &$\Gamma$~(MeV)                    &$J^{PC}$        &Process                             &experiment   \\ \hline

$X(3860)$    &$3862^{+26}_{-32}{}^{+40}_{-13}$   &$201^{+154}_{-67}{}^{+88}_{-82}$  &$0^{++}$        &$e^+ e^- \to J/\psi D\bar{D}$       &Belle \\

$X(3872)$    &$3871.69\pm0.17$                   &$<1.2$                            &$1^{++}$        &$B\to K \pi^+\pi^-J/\psi$           &Belle   \\

$X(3915)$    &$3918.4\pm1.9$                     &$20\pm$5                          &$0/2^{++}$      &$B\to KJ/\psi\omega$                &Belle, BaBar  \\

$X(3940)$    &$3942^{+7}_{-6}\pm6$               &$37^{+26}_{-15}\pm8$              &$?^{??}$        &$e^+e^-\to J/\psi D\bar{D}^*$       &Belle   \\

$X(4140)$    &$4146.8\pm2.4$                     &$22^{+8}_{-7}$                    &$1^{++}$        &$B^+\to J/\psi \phi K^+$            &CDF, D0, LHCb \\

$X(4160)$    &$4156^{+25}_{-20}\pm15 $           &$139^{+111}_{-61}\pm21$           &$?^{??}$        &$e^+e^- \to J/\psi D^*\bar{D}^*$    &Belle \\

$X(4274)$    &$4274^{+8}_{-6}$                   &$49\pm12$                         &$1^{++}$        &$B^+\to J/\psi \phi K^+$            &CDF,  LHCb \\

$X(4350)$    &$4350.6^{+4.6}_{-5.1}\pm0.7$       &$13^{+18}_{-9}\pm4$               &$?^{?+}$        &$e^+e^-\to \phi J/\psi$             &Belle \\

$X(4500)$    &$4506\pm11^{+12}_{-15}$            &$92\pm21^{+21}_{-20}$             &$0^{++}$        &$B^+\to J/\psi \phi K^+$            &LHCb \\

$X(4700)$    &$4704\pm10^{+14}_{-24}$            &$120\pm31^{+42}_{-33}$            &$0^{++}$        &$B^+\to J/\psi \phi K^+$            &LHCb \\
\hline

$Y(4220)$    &$4218^{+5}_{-4}$                   &$59^{+12}_{-10}$                  &$1^{--}$        &$e^+e^-\to h_c \pi^+ \pi^-$         &BESIII \\

$Y(4360)$    &$4368 \pm 13$                      &$96\pm7$                          &$1^{--}$        &$e^+e^-\to\pi^+\pi^-\psi^\prime$    &BaBar, Belle \\

$Y(4390)$    &$4391.5^{+6.3}_{-6.8}\pm1.0$       &$139.5^{+16.2}_{-20.6}\pm0.6$     &$1^{--}$        &$e^+e^- \to h_c \pi^+ \pi^-$        &BESIII \\

$Y(4660)$    &$4643\pm9$                         &$72\pm11$                         &$1^{--}$        &$e^+e^-\to\pi^+\pi^- \psi^\prime$   &Belle, BaBar \\ \hline

$Z_c(3900)$  &$3887.2\pm2.3$                     &$28.2\pm2.6$                      &$1^{+-}$        &$Y(4260) \to J/\psi\pi^+\pi^-$      &BESIII, Belle  \\

$Z_c(4020)$  &$4024.1\pm1.9$                     &$13\pm5$                          &$?^{?-}$        &$e^+e^-\to \pi^-\pi^+h_c$           &BESIII \\

$Z_c(4050)$  &$4051\pm14^{+20}_{-41}$            &$82^{+21}_{-17}{}^{+47}_{-22}$    &$?^{?+}$        &$\bar{B}^0\to K^-\pi^+\chi_{c1}$    &Belle \\

$Z_c(4055)$  &$4054\pm3\pm1$                     &$45\pm11\pm6$                     &$?^{?-}$        &$e^+e^- \to \pi^-\pi^+\psi^\prime$  &Belle \\

$Z_c(4100)$  &$4096\pm20{}^{+18}_{-22}$          &$152\pm58{}^{+60}_{-35}$          &$0^{++}/1^{-+}$ &$B^0\to K^+\pi^-\eta_c$             &LHCb \\

$Z_c(4200)$  &$4196^{+31}_{-29}{}^{+17}_{-13}$   &$370\pm70^{+70}_{-132}$           &$1^{+-}$        &$\bar{B}^0\to K^-\pi^+J/\psi$       &Belle \\

$Z_c(4250)$  &$4248^{+44}_{-29}{}^{+180}_{-35}$  &$177^{+54}_{-39}{}^{+316}_{-61}$  &$?^{?+}$        &$\bar{B}^0\to K^-\pi^+\chi_{c1}$    &Belle \\

$Z_c(4430)$  &$4478^{+15}_{-18}$                 &$181\pm31$                        &$1^{+-}$        &$B\to K^- \pi^+\psi^\prime$         &Belle, LHCb \\
\hline\hline
\end{tabular}
\end{center}
\caption{ The masses,  widths and $J^{PC}$ of the $X$, $Y$ and $Z$ states in the $\bar{c}c$ region from the Particle Data Group except for the $Z_c(4100)$. }\label{Exp-XYZ-Table}
\end{table}

There have seen several possible interpretations for those $X$, $Y$ and $Z$ states, such as the tetraquark states, hadronic molecular states, dynamically generated resonances,  hadroquarkonium,
kinematical effects, cusp effects,  virtual states, etc using the phenomenological approaches (potential quark models), effective field theories for QCD (such as heavy quark effective field theory, nonrelativistic QCD, potential nonrelativistic QCD, Born-Oppenheimer approximation,  chiral unitary models), QCD sum rules, lattice QCD, etc. For comprehensive reviews, one can consult Refs.\cite{HXChen-review-1601, RFLebed-review-1610,AEsposito-review-1611,FKGuo-review-1705, AAli-review-1706,SLOlsen-review-1708,MNielsen-review-1812,YRLiu-review-1903,CPShen-review-1907}. In the present work, we will focus on the tetraquark interpretations.

The QCD sum rules is a powerful theoretical approach in studying the hadron properties, and has been applied extensively to calculate the masses, decay constants, form-factors, hadronic coupling constants, etc \cite{SVZ79,Reinders85,Colangelo-Review}. In 2006, R. D. Matheus et al took the $X(3872) $ as the $J^{PC}=1^{++}$ diquark-antidiquark type  tetraquark state, and studied  its mass with the QCD sum rules   by carrying out the operator product expansion up to the vacuum condensates  of dimension 8 \cite{Narison-3872}.  Thereafter   the QCD sum rules became a powerful theoretical approach in studying the masses and widths of the $X$, $Y$ and $Z$ states, irrespective of the  hidden-charm (or hidden-bottom) tetraquark states or hadronic molecular states  \cite{MNielsen-review-1812,Narison-3872,WangHuangtao-2014-PRD,ZhangJR,Wang-tetra-formula,WangHuang-2014-NPA,Chen-Zhu,C.F.Qiao,WangZG-V-tetra,WangZG-V-Y4660,Azizi}.
In the QCD sum rules, we choose the color-antitriplet-color-triplet ($ \bar{\bf3}_c{\bf 3}_c$) type, in other words, the diquark-antidiquark type, color-sextet-color-antisextet (${\bf 6}_c\bar{\bf 6}_c$) type, color-singlet-color-singlet (${\bf 1}_c{\bf 1}_c$) type and color-octet-color-octet (${\bf 8}_c{\bf 8}_c$) type local four-quark currents to study the tetraquark states. It is better to call the corresponding tetraquark states as the $ \bar{\bf3}_c{\bf 3}_c$-type, ${\bf 6}_c\bar{\bf 6}_c$-type, ${\bf 1}_c{\bf 1}_c$-type and $ {\bf8}_c{\bf 8}_c$-type tetraquark states, respectively. In the literatures, we usually call the $ \bar{\bf3}_c{\bf 3}_c$-type and ${\bf 1}_c{\bf 1}_c$-type tetraquark states as the tetraquark states and (tetraquark or hadronic) molecular states, respectively. Thereafter, we will use the name ${\bf 1}_c{\bf 1}_c$-type tetraquark states in stead of the name (tetraquark or hadronic) molecular states according to the local currents.

 In the QCD sum rules for the hidden-charm (or hidden-bottom) tetraquark states and ${\bf1}_c{\bf 1}_c$-type tetraquark states, the integrals
 \begin{eqnarray}
 \int_{4m_Q^2(\mu)}^{s_0} ds \,\rho_{QCD}(s,\mu)\exp\left(-\frac{s}{T^2} \right)\, ,
 \end{eqnarray}
are sensitive to the heavy quark masses $m_Q(\mu)$, more precisely speaking, the integrals are sensitive to the energy scales $\mu$, where the $\rho_{QCD}(s,\mu)$ are the QCD spectral densities,  the $T^2$ are the Borel parameters, and the $s_0$ are the continuum threshold parameters.
In Ref.\cite{WangHuangtao-2014-PRD}, we tentatively assign the $X(3872)$ and $Z_c(3900)$ to be   the diquark-antidiquark type  axialvector tetraquark states, and study them   with the QCD sum rules in details, and  explore the energy scale dependence of the QCD sum rules for the hidden-charm   tetraquark states   for the first time \cite{WangHuangtao-2014-PRD}. In Ref.\cite{Wang-tetra-formula}, we suggest a  formula,
\begin{eqnarray}
\mu&=&\sqrt{M^2_{X/Y/Z}-(2{\mathbb{M}}_c)^2} \, ,
 \end{eqnarray}
 with the effective heavy  mass ${\mathbb{M}}_c$ to determine the optimal energy scales, which can be applied to study the
 hidden-bottom tetraquark states directly with the simple replacement ${\mathbb{M}}_c \to {\mathbb{M}}_b$ \cite{WangHuang-2014-NPA}.

In the  scenario of tetraquark  states,  the $Y$ states, i.e. the $Y(4220)$, $Y(4360)$, $Y(4390)$, $Y(4660)$, can be assigned to be  the diquark-antidiquark type tetraquark states. In Ref.\cite{WangZG-V-tetra}, we  introduce a relative P-wave between the diquark and antidiquark operators  explicitly in constructing  the  tetraquark currents to study the  vector tetraquark states  with the QCD sum rules systematically, and  obtain the lowest vector tetraquark masses up to now, which support assigning the $Y(4220/4260)$,  $Y(4320/4360)$, $Y(4390)$ and $Z_c(4250)$ to be the vector hidden-charm tetraquark   states. While novel analysis of the masses and widths of the
 vector hidden-charm tetraquark states without a relative P-wave between the diquark and antidiquark constituents indicate that the $Y(4660)$ can be assigned to be
 a $[sc]_P[\bar{s}\bar{c}]_A-[sc]_A[\bar{s}\bar{c}]_P$  type tetraquark state \cite{WangZG-V-Y4660}. In those studies, the energy scale formula and modified energy scale formula play an important role in enhancing the pole contributions and in improving the convergent behavior of the operator product expansion.

 The $X$ states $X(4140)$, $X(4274)$, $X(4350)$, $X(4500)$ and $X(4700)$ are observed in the $J/\psi\phi$ mass spectrum, if their dominant Fock components are tetraquark states, their quark constituents are $\bar{c}c\bar{s}s$ rather than $\bar{c}c\bar{q}q$. The QCD sum rules support assigning the $X(3915)$, $X(4140)$, $X(4274)$, $X(4500)$ and $X(4700)$ to be the diquark-antidiquark type tetraquark states \cite{WangZG-X4500,WangZG-Di-Y4140}, the decay $X(3915)\to J/\psi\phi \to J/\psi\omega$ can take place through the $\phi-\omega$ mixing \cite{Lebed-X3915}.

The $X$ states $X(3860)$, $X(3872)$, $X(3940)$ and $X(4160)$ are observed in the final states $D\bar{D}$, $D\bar{D}^*$, $D^*\bar{D}$,  $D^*\bar{D}^*$ or $J/\psi \pi\pi$, if their dominant Fock components  are tetraquark states, their constituents are $\bar{c}c\bar{q}q$.

The $Z$ states $Z_c(3900)$, $Z_c(4020)$, $Z_c(4050)$, $Z_c(4055)$, $Z_c(4100)$, $Z_c(4200)$, $Z_c(4250)$ and $Z_c(4430)$ have non-zero electric charge, which prevent them from being the conventional two-quark mesons, they are excellent candidates for the tetraquark states. Those $Z_c$ states have been studied with the QCD sum rules in one way or the other \cite{MNielsen-review-1812,WangHuangtao-2014-PRD,Wang-tetra-formula,C.F.Qiao,Azizi,WangZG-Z4100-1903,ChenHX-Z4600-A,WangZG-axial-Z4600}.

We usually take the diquarks in color antitriplet ${\bar {\bf 3}}_c$ as the basic building blocks to construct  the tetraquark states.   The diquarks  operators $\varepsilon^{abc}q^{T}_b C\Gamma q^{\prime}_c$  have  five  structures  in Dirac spinor space, where $C\Gamma=C\gamma_5$, $C$, $C\gamma_\mu \gamma_5$,  $C\gamma_\mu $ and $C\sigma_{\mu\nu}$ (or $C\sigma_{\mu\nu}\gamma_5$) for the scalar ($S$), pseudoscalar ($P$), vector ($V$), axialvector ($A$)  and  tensor ($T$) diquarks, respectively, the $a$, $b$, $c$ are color indexes. The tensor diquark states have both $J^P=1^+$ and $1^-$ components, we project out the $1^+$ and $1^-$ components explicitly, and denote the corresponding axialvector and vector diquarks as $\widetilde{A}$ and $\widetilde{V}$, respectively.

All in all, those $X$, $Y$ and $Z$ states have been studied with the QCD sum rules in one way or the other \cite{MNielsen-review-1812,Narison-3872,WangHuangtao-2014-PRD,ZhangJR,Wang-tetra-formula,WangHuang-2014-NPA,Chen-Zhu,C.F.Qiao,WangZG-V-tetra,WangZG-V-Y4660,Azizi,WangZG-X4500,
WangZG-Di-Y4140,WangZG-Z4100-1903,ChenHX-Z4600-A,WangZG-axial-Z4600},
in  the present work,  we take the scalar, pseudoscalar, axialvector, vector and tensor   diquark operators  as the basic building blocks    to construct twenty  tetraquark currents,  and study the mass spectrum of the hidden-charm tetraquark states with the QCD sum rules in a comprehensive way,  and revisit the assignments of the  $X$ and $Z$ states in the  scenario of tetraquark  states  and try to accommodate the exotic states as many as possible in a consistent way.
We take the energy scale formula  $\mu=\sqrt{M^2_{X/Y/Z}-(2{\mathbb{M}}_c)^2}$ to determine the best energy scales of the QCD spectral densities so as to enhance  the pole contributions and improve the convergent behaviors of the operator product expansion \cite{Wang-tetra-formula}. In other words, the predicted hidden-charm tetraquark masses and the QCD spectral densities satisfy  the relation $M^2_{X/Y/Z}=\mu^2+4{\mathbb{M}}_c^2$, where the ${\mathbb{M}}_c$ has an universal value. It is an unique feature of our works.
We obtain new results for eleven tetraquark currents and present them here for the first time, for  other tetraquark currents, we recalculate the vacuum condensates up to dimension $10$ in the operator product expansion consistently and preform updated  analysis \cite{WangHuangtao-2014-PRD,WangZG-Z4100-1903,WangZG-axial-Z4600}.
Those hidden-charm tetraquark states may be observed  at the BESIII, LHCb, Belle II,  CEPC, FCC, ILC  in the future, and shed light on the nature of the exotic $X$, $Y$, $Z$ states.

The article is arranged as follows:  we derive the QCD sum rules for the masses and pole residues  of  the  hidden-charm tetraquark states in section 2; in section 3, we   present the numerical results and discussions; section 4 is reserved for our conclusion.

\section{QCD sum rules for  the  hidden-charm  tetraquark states}

We write down  the two-point correlation functions $\Pi(p)$, $\Pi_{\mu\nu}(p)$ and $\Pi_{\mu\nu\alpha\beta}(p)$  in the QCD sum rules,
\begin{eqnarray}\label{CF-Pi}
\Pi(p)&=&i\int d^4x e^{ip \cdot x} \langle0|T\Big\{J(x)J^{\dagger}(0)\Big\}|0\rangle \, ,\nonumber\\
\Pi_{\mu\nu}(p)&=&i\int d^4x e^{ip \cdot x} \langle0|T\Big\{J_\mu(x)J_{\nu}^{\dagger}(0)\Big\}|0\rangle \, ,\nonumber\\
\Pi_{\mu\nu\alpha\beta}(p)&=&i\int d^4x e^{ip \cdot x} \langle0|T\Big\{J_{\mu\nu}(x)J_{\alpha\beta}^{\dagger}(0)\Big\}|0\rangle \, ,
\end{eqnarray}
where the currents $J(x)=J_{SS}(x)$,  $J_{AA}(x)$, $J_{\widetilde{A}\widetilde{A}}(x)$, $J_{VV}(x)$,
$J_{\widetilde{V}\widetilde{V}}(x)$, $J_{PP}(x)$, $J_\mu(x)=J^{SA}_{-,\mu}(x)$, $J_{-,\mu}^{\widetilde{A}A}(x)$, $J_{-,\mu}^{\widetilde{V}V}(x)$, $J^{PV}_{-,\mu}(x)$, $J^{SA}_{+,\mu}(x)$,
$J_{+,\mu}^{\widetilde{V}V}(x)$, $J_{+,\mu}^{\widetilde{A}A}(x)$, $J^{PV}_{+,\mu}(x)$,
$J_{\mu\nu}(x)=J^{AA}_{-,\mu\nu}(x)$, $J^{S\widetilde{A}}_{-,\mu\nu}(x)$, $J^{VV}_{-,\mu\nu}(x)$, $J^{S\widetilde{A}}_{+,\mu\nu}(x)$, $J^{AA}_{+,\mu\nu}(x)$, $J^{VV}_{+,\mu\nu}(x)$,
\begin{eqnarray}
J_{SS}(x)&=&\varepsilon^{ijk}\varepsilon^{imn}u^{Tj}(x)C\gamma_5 c^k(x)  \bar{d}^m(x)\gamma_5 C \bar{c}^{Tn}(x) \, ,\nonumber \\
J_{AA}(x)&=&\varepsilon^{ijk}\varepsilon^{imn}u^{Tj}(x)C\gamma_\mu c^k(x)  \bar{d}^m(x)\gamma^\mu C \bar{c}^{Tn}(x) \, ,\nonumber \\
J_{\tilde{A}\tilde{A}}(x)&=&\varepsilon^{ijk}\varepsilon^{imn}u^{Tj}(x)C\sigma^v_{\mu\nu} c^k(x)  \bar{d}^m(x)\sigma_v^{\mu\nu} C \bar{c}^{Tn}(x) \, ,\nonumber \\
J_{VV}(x)&=&\varepsilon^{ijk}\varepsilon^{imn}u^{Tj}(x)C\gamma_\mu\gamma_5 c^k(x)  \bar{d}^m(x)\gamma_5\gamma^\mu C \bar{c}^{Tn}(x) \, ,\nonumber \\
J_{\tilde{V}\tilde{V}}(x)&=&\varepsilon^{ijk}\varepsilon^{imn}u^{Tj}(x)C\sigma^t_{\mu\nu} c^k(x)  \bar{d}^m(x)\sigma_t^{\mu\nu} C \bar{c}^{Tn}(x) \, ,\nonumber \\
J_{PP}(x)&=&\varepsilon^{ijk}\varepsilon^{imn}u^{Tj}(x)Cc^k(x)  \bar{d}^m(x) C \bar{c}^{Tn}(x) \, ,
\end{eqnarray}

\begin{eqnarray}
J^{SA}_{-,\mu}(x)&=&\frac{\varepsilon^{ijk}\varepsilon^{imn}}{\sqrt{2}}\Big[u^{Tj}(x)C\gamma_5c^k(x) \bar{d}^m(x)\gamma_\mu C \bar{c}^{Tn}(x)-u^{Tj}(x)C\gamma_\mu c^k(x)\bar{d}^m(x)\gamma_5C \bar{c}^{Tn}(x) \Big] \, ,\nonumber\\
J^{AA}_{-,\mu\nu}(x)&=&\frac{\varepsilon^{ijk}\varepsilon^{imn}}{\sqrt{2}}\Big[u^{Tj}(x) C\gamma_\mu c^k(x) \bar{d}^m(x) \gamma_\nu C \bar{c}^{Tn}(x)  -u^{Tj}(x) C\gamma_\nu c^k(x) \bar{d}^m(x) \gamma_\mu C \bar{c}^{Tn}(x) \Big] \, , \nonumber\\
J^{S\widetilde{A}}_{-,\mu\nu}(x)&=&\frac{\varepsilon^{ijk}\varepsilon^{imn}}{\sqrt{2}}\Big[u^{Tj}(x)C\gamma_5 c^k(x)  \bar{d}^m(x)\sigma_{\mu\nu} C \bar{c}^{Tn}(x)- u^{Tj}(x)C\sigma_{\mu\nu} c^k(x)  \bar{d}^m(x)\gamma_5 C \bar{c}^{Tn}(x) \Big] \, , \nonumber\\
J_{-,\mu}^{\widetilde{A}A}(x)&=&\frac{\varepsilon^{ijk}\varepsilon^{imn}}{\sqrt{2}}\Big[u^{Tj}(x)C\sigma_{\mu\nu}\gamma_5 c^k(x)\bar{d}^m(x)\gamma^\nu C \bar{c}^{Tn}(x)-u^{Tj}(x)C\gamma^\nu c^k(x)\bar{d}^m(x)\gamma_5\sigma_{\mu\nu} C \bar{c}^{Tn}(x) \Big] \, , \nonumber\\
J_{-,\mu}^{\widetilde{V}V}(x)&=&\frac{\varepsilon^{ijk}\varepsilon^{imn}}{\sqrt{2}}\left[u^{Tj}(x)C\sigma_{\mu\nu} c^k(x)\bar{d}^m(x)\gamma_5\gamma^\nu C \bar{c}^{Tn}(x)+u^{Tj}(x)C\gamma^\nu \gamma_5c^k(x)\bar{d}^m(x) \sigma_{\mu\nu} C \bar{c}^{Tn}(x) \right] \, , \nonumber\\
J^{VV}_{-,\mu\nu}(x)&=&\frac{\varepsilon^{ijk}\varepsilon^{imn}}{\sqrt{2}}\Big[u^{Tj}(x) C\gamma_\mu \gamma_5c^k(x) \bar{d}^m(x) \gamma_5\gamma_\nu C \bar{c}^{Tn}(x)  -u^{Tj}(x) C\gamma_\nu\gamma_5 c^k(x) \bar{d}^m(x) \gamma_5\gamma_\mu C \bar{c}^{Tn}(x) \Big] \, , \nonumber\\
J^{PV}_{-,\mu}(x)&=&\frac{\varepsilon^{ijk}\varepsilon^{imn}}{\sqrt{2}}\Big[u^{Tj}(x)Cc^k(x) \bar{d}^m(x)\gamma_5\gamma_\mu C \bar{c}^{Tn}(x)+u^{Tj}(x)C\gamma_\mu \gamma_5c^k(x)\bar{d}^m(x)C \bar{c}^{Tn}(x) \Big] \, ,
\end{eqnarray}

\begin{eqnarray}
J^{SA}_{+,\mu}(x)&=&\frac{\varepsilon^{ijk}\varepsilon^{imn}}{\sqrt{2}}\Big[u^{Tj}(x)C\gamma_5c^k(x) \bar{d}^m(x)\gamma_\mu C \bar{c}^{Tn}(x)+u^{Tj}(x)C\gamma_\mu c^k(x)\bar{d}^m(x)\gamma_5C \bar{c}^{Tn}(x) \Big] \, ,\nonumber\\
J^{S\widetilde{A}}_{+,\mu\nu}(x)&=&\frac{\varepsilon^{ijk}\varepsilon^{imn}}{\sqrt{2}}\Big[u^{Tj}(x)C\gamma_5 c^k(x)  \bar{d}^m(x)\sigma_{\mu\nu} C \bar{c}^{Tn}(x)+ u^{Tj}(x)C\sigma_{\mu\nu} c^k(x)  \bar{d}^m(x)\gamma_5 C \bar{c}^{Tn}(x) \Big] \, , \nonumber\\
J_{+,\mu}^{\widetilde{V}V}(x)&=&\frac{\varepsilon^{ijk}\varepsilon^{imn}}{\sqrt{2}}\left[u^{Tj}(x)C\sigma_{\mu\nu} c^k(x)\bar{d}^m(x)\gamma_5\gamma^\nu C \bar{c}^{Tn}(x)-u^{Tj}(x)C\gamma^\nu \gamma_5c^k(x)\bar{d}^m(x) \sigma_{\mu\nu} C \bar{c}^{Tn}(x) \right] \, , \nonumber\\
J_{+,\mu}^{\widetilde{A}A}(x)&=&\frac{\varepsilon^{ijk}\varepsilon^{imn}}{\sqrt{2}}\left[u^{Tj}(x)C\sigma_{\mu\nu}\gamma_5 c^k(x)\bar{d}^m(x)\gamma^\nu C \bar{c}^{Tn}(x)+u^{Tj}(x)C\gamma^\nu c^k(x)\bar{d}^m(x)\gamma_5\sigma_{\mu\nu} C \bar{c}^{Tn}(x) \right] \, , \nonumber\\
J^{PV}_{+,\mu}(x)&=&\frac{\varepsilon^{ijk}\varepsilon^{imn}}{\sqrt{2}}\Big[u^{Tj}(x)Cc^k(x) \bar{d}^m(x)\gamma_5\gamma_\mu C \bar{c}^{Tn}(x)-u^{Tj}(x)C\gamma_\mu \gamma_5c^k(x)\bar{d}^m(x)C \bar{c}^{Tn}(x) \Big] \, ,\nonumber\\
J^{AA}_{+,\mu\nu}(x)&=&\frac{\varepsilon^{ijk}\varepsilon^{imn}}{\sqrt{2}}\Big[u^{Tj}(x) C\gamma_\mu c^k(x) \bar{d}^m(x) \gamma_\nu C \bar{c}^{Tn}(x)  +u^{Tj}(x) C\gamma_\nu c^k(x) \bar{d}^m(x) \gamma_\mu C \bar{c}^{Tn}(x) \Big] \, , \nonumber\\
J^{VV}_{+,\mu\nu}(x)&=&\frac{\varepsilon^{ijk}\varepsilon^{imn}}{\sqrt{2}}\Big[u^{Tj}(x) C\gamma_\mu \gamma_5c^k(x) \bar{d}^m(x) \gamma_5\gamma_\nu C \bar{c}^{Tn}(x)  +u^{Tj}(x) C\gamma_\nu\gamma_5 c^k(x) \bar{d}^m(x) \gamma_5\gamma_\mu C \bar{c}^{Tn}(x) \Big] \, , \nonumber\\
\end{eqnarray}
$\sigma^t_{\mu\nu} =\frac{i}{2}\Big[\gamma^t_\mu, \gamma^t_\nu \Big]$, $\sigma^v_{\mu\nu} =\frac{i}{2}\Big[\gamma^v_\mu, \gamma^t_\nu \Big]$,
$\gamma^v_\mu =  \gamma \cdot t t_\mu$, $\gamma^t_\mu=\gamma_\mu-\gamma \cdot t t_\mu$, $t^\mu=(1,\vec{0})$,  the $i$, $j$, $k$, $m$, $n$ are  color indexes,
    the $C$ is the charge conjugation matrix, the subscripts $\pm$ denote the positive charge  conjugation and negative charge conjugation, respectively, the superscripts or subscripts $P$, $S$, $A$($\widetilde{A}$) and $V$($\widetilde{V}$) denote the pseudoscalar, scalar, axialvector and vector diquark and antidiquark operators, respectively \cite{WangHuangtao-2014-PRD,WangHuang-2014-NPA,WangZG-X4500,WangZG-Di-Y4140,WangZG-Z4100-1903,WangZG-axial-Z4600}.
For the currents  $J_{SS}(x)$,  $J_{\widetilde{A}\widetilde{A}}(x)$, $J_{\widetilde{V}\widetilde{V}}(x)$, $J^{SA}_{-,\mu}(x)$,
$J^{AA}_{-,\mu\nu}(x)$, $J_{-,\mu}^{\widetilde{A}A}(x)$, $J_{-,\mu}^{\widetilde{V}V}(x)$, $J^{SA}_{+,\mu}(x)$ and $J^{AA}_{+,\mu\nu}(x)$, we update the old calculations.   For the currents $J_{AA}(x)$, $J_{VV}(x)$, $J_{PP}(x)$, $J^{S\widetilde{A}}_{-,\mu\nu}(x)$, $J^{VV}_{-,\mu\nu}(x)$,
$J^{PV}_{-,\mu}(x)$, $J^{S\widetilde{A}}_{+,\mu\nu}(x)$, $J_{+,\mu}^{\widetilde{V}V}(x)$,
$J_{+,\mu}^{\widetilde{A}A}(x)$, $J^{PV}_{+,\mu}(x)$ and
$J^{VV}_{+,\mu\nu}(x)$, we obtain new results.
  Under parity transform $\widehat{P}$, the current  operators have the  properties,
\begin{eqnarray}
\widehat{P} J(x)\widehat{P}^{-1}&=&+J(\tilde{x}) \, , \nonumber\\
\widehat{P} J_\mu(x)\widehat{P}^{-1}&=&-J^\mu(\tilde{x}) \, , \nonumber\\
\widehat{P} J^{S\widetilde{A}}_{\mu\nu}(x)\widehat{P}^{-1}&=&-J_{S\widetilde{A}}^{\mu\nu}(\tilde{x}) \, , \nonumber\\
\widehat{P} J^{AA/VV}_{\mu\nu}(x)\widehat{P}^{-1}&=&+J_{AA/VV}^{\mu\nu}(\tilde{x}) \, ,
\end{eqnarray}
where  $x^\mu=(t,\vec{x})$ and $\tilde{x}^\mu=(t,-\vec{x})$, and we have neglected other superscripts and subscripts of the current operators.

 The current operators $J(x)$, $J_\mu(x)$ and $J_{\mu\nu}(x)$ have the  symbolic quark structure  $\bar{c}c\bar{d}u$ with the isospin $I=1$ and $I_3=1$,
 other currents in the isospin multiplets can be constructed analogously,  for example, we can write down the corresponding isospin singlet current for  the $J_{SS}(x)$ directly,
\begin{eqnarray}
J^{I=0}_{SS}(x)&=&\frac{\varepsilon^{ijk}\varepsilon^{imn}}{\sqrt{2}}\Big[u^{Tj}(x)C\gamma_5 c^k(x)  \bar{u}^m(x)\gamma_5 C \bar{c}^{Tn}(x)+d^{Tj}(x)C\gamma_5 c^k(x)  \bar{d}^m(x)\gamma_5 C \bar{c}^{Tn}(x)\Big] \, . \nonumber \\
\end{eqnarray}
 In the isospin limit, the current operators with the  symbolic quark structures $\bar{c}c\bar{d}u$, $\bar{c}c\bar{u}d$, $\bar{c}c\frac{\bar{u}u-\bar{d}d}{\sqrt{2}}$, $\bar{c}c\frac{\bar{u}u+\bar{d}d}{\sqrt{2}}$ couple potentially  to the hidden-charm
tetraquark states with degenerated  masses, the current operators with $I=1$ and $0$ lead to the same QCD sum rules.  Thereafter, we will denote the $Z_c$ states as the isospin triplet, and the $X$ states as the isospin singlet,
\begin{eqnarray}
Z_c&:&\bar{c}c\bar{d}u\, ,\, \bar{c}c\bar{u}d\, ,\, \bar{c}c\frac{\bar{u}u-\bar{d}d}{\sqrt{2}}\, , \nonumber\\
X&:& \bar{c}c\frac{\bar{u}u+\bar{d}d}{\sqrt{2}}\, .
\end{eqnarray}

The current operators with the symbolic quark structures $\bar{c}c\frac{\bar{u}u-\bar{d}d}{\sqrt{2}}$ and $\bar{c}c\frac{\bar{u}u+\bar{d}d}{\sqrt{2}}$ have definite
charge conjugation. In this article, we will assume  that the $\bar{c}c\bar{d}u$ type tetraquark states have the same charge conjugation as their neutral charge partners.

  Under charge conjugation transform $\widehat{C}$, the currents $J(x)$, $J_\mu(x)$ and $J_{\mu\nu}(x)$ have the properties,
\begin{eqnarray}
\widehat{C}J(x)\widehat{C}^{-1}&=&+ J(x)\mid_{u\leftrightarrow d} \, , \nonumber\\
\widehat{C}J_{\pm,\mu}(x)\widehat{C}^{-1}&=&\pm J_{\pm,\mu}(x)\mid_{u\leftrightarrow d}  \, , \nonumber\\
\widehat{C}J_{\pm,\mu\nu}(x)\widehat{C}^{-1}&=&\pm J_{\pm,\mu\nu}(x)\mid_{u\leftrightarrow d}  \, ,
\end{eqnarray}
where we have neglected other superscripts and subscripts of the current operators.

\begin{table}
\begin{center}
\begin{tabular}{|c|c|c|c|c|c|c|c|c|}\hline\hline
 $Z_c$                                                                            & $J^{PC}$  & Currents              \\ \hline

$[uc]_{S}[\overline{dc}]_{S}$                                                     & $0^{++}$  & $J_{SS}(x)$              \\

$[uc]_{A}[\overline{dc}]_{A}$                                                     & $0^{++}$  & $J_{AA}(x)$               \\

$[uc]_{\tilde{A}}[\overline{dc}]_{\tilde{A}}$                                     & $0^{++}$  & $J_{\widetilde{A}\widetilde{A}}(x)$             \\

$[uc]_{V}[\overline{dc}]_{V}$                                                     & $0^{++}$  & $J_{VV}(x)$               \\

$[uc]_{\tilde{V}}[\overline{dc}]_{\tilde{V}}$                                     & $0^{++}$  & $J_{\widetilde{V}\widetilde{V}}(x)$           \\

$[uc]_{P}[\overline{dc}]_{P}$                                                     & $0^{++}$  & $J_{PP}(x)$              \\ \hline

$[uc]_S[\overline{dc}]_{A}-[uc]_{A}[\overline{dc}]_S$                             & $1^{+-}$  & $J^{SA}_{-,\mu}(x)$         \\

$[uc]_{A}[\overline{dc}]_{A}$                                                     & $1^{+-}$  & $J^{AA}_{-,\mu\nu}(x)$        \\

$[uc]_S[\overline{dc}]_{\widetilde{A}}-[uc]_{\widetilde{A}}[\overline{dc}]_S$     & $1^{+-}$  & $J^{S\widetilde{A}}_{-,\mu\nu}(x)$     \\

$[uc]_{\widetilde{A}}[\overline{dc}]_{A}-[uc]_{A}[\overline{dc}]_{\widetilde{A}}$ & $1^{+-}$  & $J_{-,\mu}^{\widetilde{A}A}(x)$   \\

$[uc]_{\widetilde{V}}[\overline{dc}]_{V}+[uc]_{V}[\overline{dc}]_{\widetilde{V}}$ & $1^{+-}$  & $J_{-,\mu}^{\widetilde{V}V}(x)$      \\

$[uc]_{V}[\overline{dc}]_{V}$                                                     & $1^{+-}$  & $J^{VV}_{-,\mu\nu}(x)$        \\

$[uc]_P[\overline{dc}]_{V}+[uc]_{V}[\overline{dc}]_P$                             & $1^{+-}$  & $J^{PV}_{-,\mu}(x)$         \\
\hline

$[uc]_S[\overline{dc}]_{A}+[uc]_{A}[\overline{dc}]_S$                             & $1^{++}$  & $J^{SA}_{+,\mu}(x)$        \\

$[uc]_S[\overline{dc}]_{\widetilde{A}}+[uc]_{\widetilde{A}}[\overline{dc}]_S$     & $1^{++}$  & $J^{S\widetilde{A}}_{+,\mu\nu}(x)$     \\

$[uc]_{\widetilde{V}}[\overline{dc}]_{V}-[uc]_{V}[\overline{dc}]_{\widetilde{V}}$ & $1^{++}$  & $J_{+,\mu}^{\widetilde{V}V}(x)$      \\

$[uc]_{\widetilde{A}}[\overline{dc}]_{A}+[uc]_{A}[\overline{dc}]_{\widetilde{A}}$ & $1^{++}$  & $J_{+,\mu}^{\widetilde{A}A}(x)$       \\
$[uc]_P[\overline{dc}]_{V}-[uc]_{V}[\overline{dc}]_P$                             & $1^{++}$  & $J^{PV}_{+,\mu}(x)$         \\ \hline

$[uc]_{A}[\overline{dc}]_{A}$                                                     & $2^{++}$  & $J^{AA}_{+,\mu\nu}(x)$       \\

$[uc]_{V}[\overline{dc}]_{V}$                                                     & $2^{++}$  & $J^{VV}_{+,\mu\nu}(x)$        \\
\hline\hline
\end{tabular}
\end{center}
\caption{ The quark structures and corresponding current operators  for the hidden-charm tetraquark states. }\label{Current-Table}
\end{table}

At the hadron side, we  insert  a complete set of intermediate hadronic states with
the same quantum numbers as the current operators $J(x)$, $J_\mu(x)$ and $J_{\mu\nu}(x)$ into the
correlation functions $\Pi(p)$, $\Pi_{\mu\nu}(p)$ and $\Pi_{\mu\nu\alpha\beta}(p)$   to obtain the hadronic representation
\cite{SVZ79,Reinders85}, and isolate the ground state hidden-charm tetraquark contributions,
\begin{eqnarray}
\Pi(p)&=&\frac{\lambda_{Z^+}^2}{M_{Z^+}^2-p^2} +\cdots \nonumber\\
&=&\Pi_{+}(p^2) \, ,\nonumber\\
\Pi_{\mu\nu}(p)&=&\frac{\lambda_{Z^+}^2}{M_{Z^+}^2-p^2}\left( -g_{\mu\nu}+\frac{p_{\mu}p_{\nu}}{p^2}\right) +\cdots \nonumber\\
&=&\Pi_{+}(p^2)\left( -g_{\mu\nu}+\frac{p_{\mu}p_{\nu}}{p^2}\right)+\cdots \, ,\nonumber\\
\Pi^{AA,-}_{\mu\nu\alpha\beta}(p)&=&\frac{\lambda_{ Z^+}^2}{M_{Z^+}^2\left(M_{Z^+}^2-p^2\right)}\left(p^2g_{\mu\alpha}g_{\nu\beta} -p^2g_{\mu\beta}g_{\nu\alpha} -g_{\mu\alpha}p_{\nu}p_{\beta}-g_{\nu\beta}p_{\mu}p_{\alpha}+g_{\mu\beta}p_{\nu}p_{\alpha}+g_{\nu\alpha}p_{\mu}p_{\beta}\right) \nonumber\\
&&+\frac{\lambda_{ Z^-}^2}{M_{Z^-}^2\left(M_{Z^-}^2-p^2\right)}\left( -g_{\mu\alpha}p_{\nu}p_{\beta}-g_{\nu\beta}p_{\mu}p_{\alpha}+g_{\mu\beta}p_{\nu}p_{\alpha}+g_{\nu\alpha}p_{\mu}p_{\beta}\right) +\cdots  \nonumber\\
&=&\widetilde{\Pi}_{+}(p^2)\left(p^2g_{\mu\alpha}g_{\nu\beta} -p^2g_{\mu\beta}g_{\nu\alpha} -g_{\mu\alpha}p_{\nu}p_{\beta}-g_{\nu\beta}p_{\mu}p_{\alpha}+g_{\mu\beta}p_{\nu}p_{\alpha}+g_{\nu\alpha}p_{\mu}p_{\beta}\right) \nonumber\\
&&+\widetilde{\Pi}_{-}(p^2)\left( -g_{\mu\alpha}p_{\nu}p_{\beta}-g_{\nu\beta}p_{\mu}p_{\alpha}+g_{\mu\beta}p_{\nu}p_{\alpha}+g_{\nu\alpha}p_{\mu}p_{\beta}\right) \, ,\nonumber\\
\Pi^{S\widetilde{A},\pm}_{\mu\nu\alpha\beta}(p)&=&\frac{\lambda_{ Z^-}^2}{M_{Z^-}^2\left(M_{Z^-}^2-p^2\right)}\left(p^2g_{\mu\alpha}g_{\nu\beta} -p^2g_{\mu\beta}g_{\nu\alpha} -g_{\mu\alpha}p_{\nu}p_{\beta}-g_{\nu\beta}p_{\mu}p_{\alpha}+g_{\mu\beta}p_{\nu}p_{\alpha}+g_{\nu\alpha}p_{\mu}p_{\beta}\right) \nonumber\\
&&+\frac{\lambda_{ Z^+}^2}{M_{Z^+}^2\left(M_{Z^+}^2-p^2\right)}\left( -g_{\mu\alpha}p_{\nu}p_{\beta}-g_{\nu\beta}p_{\mu}p_{\alpha}+g_{\mu\beta}p_{\nu}p_{\alpha}+g_{\nu\alpha}p_{\mu}p_{\beta}\right) +\cdots  \nonumber\\
&=&\widetilde{\Pi}_{-}(p^2)\left(p^2g_{\mu\alpha}g_{\nu\beta} -p^2g_{\mu\beta}g_{\nu\alpha} -g_{\mu\alpha}p_{\nu}p_{\beta}-g_{\nu\beta}p_{\mu}p_{\alpha}+g_{\mu\beta}p_{\nu}p_{\alpha}+g_{\nu\alpha}p_{\mu}p_{\beta}\right) \nonumber\\
&&+\widetilde{\Pi}_{+}(p^2)\left( -g_{\mu\alpha}p_{\nu}p_{\beta}-g_{\nu\beta}p_{\mu}p_{\alpha}+g_{\mu\beta}p_{\nu}p_{\alpha}+g_{\nu\alpha}p_{\mu}p_{\beta}\right) \, , \nonumber\\
\Pi_{\mu\nu\alpha\beta}^{AA,+}(p)&=&\frac{\lambda_{ Z^+}^2}{M_{Z^+}^2-p^2}\left( \frac{\widetilde{g}_{\mu\alpha}\widetilde{g}_{\nu\beta}+\widetilde{g}_{\mu\beta}\widetilde{g}_{\nu\alpha}}{2}-\frac{\widetilde{g}_{\mu\nu}\widetilde{g}_{\alpha\beta}}{3}\right) +\cdots \, \, , \nonumber \\
&=&\Pi_{+}(p^2)\left( \frac{\widetilde{g}_{\mu\alpha}\widetilde{g}_{\nu\beta}+\widetilde{g}_{\mu\beta}\widetilde{g}_{\nu\alpha}}{2}-\frac{\widetilde{g}_{\mu\nu}\widetilde{g}_{\alpha\beta}}{3}\right) +\cdots\, ,
\end{eqnarray}
where $\widetilde{g}_{\mu\nu}=g_{\mu\nu}-\frac{p_{\mu}p_{\nu}}{p^2}$.  We add the superscripts $\pm$ in the correlation functions $\Pi^{AA,-}_{\mu\nu\alpha\beta}(p)$, $\Pi^{S\widetilde{A},\pm}_{\mu\nu\alpha\beta}(p)$ and $\Pi_{\mu\nu\alpha\beta}^{AA,+}(p)$ to denote the positive and negative charge conjugation, respectively,
  add the superscripts (subscripts) $\pm$ in the hidden-charm tetraquark states $Z_c^{\pm}$ (the $\Pi_{\pm}(p^2)$ and $\widetilde{\Pi}_{\pm}(p^2)$ components of the correlation functions) to denote the positive  and negative parity contributions, respectively. The correlation functions $\Pi^{VV,\pm}_{\mu\nu\alpha\beta}(p)$ and $\Pi^{AA,\pm}_{\mu\nu\alpha\beta}(p)$ have the same tensor structures, we neglect the explicit expressions of the $\Pi^{VV,\pm}_{\mu\nu\alpha\beta}(p)$ for simplicity.  The   pole residues  or current-tetraquark coupling constants $\lambda_{Z^\pm}$ are defined by
\begin{eqnarray}
 \langle 0|J(0)|Z_c^+(p)\rangle &=&\lambda_{Z^+}\, , \nonumber\\
 \langle 0|J_\mu(0)|Z_c^+(p)\rangle &=&\lambda_{Z^+}\varepsilon_\mu\, , \nonumber\\
  \langle 0|J_{\pm,\mu\nu}^{S\widetilde{A}}(0)|Z_c^-(p)\rangle &=& \frac{\lambda_{Z^-}}{M_{Z^-}} \, \varepsilon_{\mu\nu\alpha\beta} \, \varepsilon^{\alpha}p^{\beta}\, , \nonumber\\
 \langle 0|J_{\pm,\mu\nu}^{S\widetilde{A}}(0)|Z_c^+(p)\rangle &=&\frac{\lambda_{Z^+}}{M_{Z^+}} \left(\varepsilon_{\mu}p_{\nu}-\varepsilon_{\nu}p_{\mu} \right)\, , \nonumber\\
  \langle 0|J_{-,\mu\nu}^{AA/VV}(0)|Z_c^+(p)\rangle &=& \frac{\lambda_{Z^+}}{M_{Z^+}} \, \varepsilon_{\mu\nu\alpha\beta} \, \varepsilon^{\alpha}p^{\beta}\, , \nonumber\\
 \langle 0|J_{-,\mu\nu}^{AA/VV}(0)|Z_c^-(p)\rangle &=&\frac{\lambda_{Z^-}}{M_{Z^-}} \left(\varepsilon_{\mu}p_{\nu}-\varepsilon_{\nu}p_{\mu} \right)\, , \nonumber\\
  \langle 0|J_{+,\mu\nu}^{AA/VV}(0)|Z_c^+(p)\rangle &=& \lambda_{Z^+}\, \varepsilon_{\mu\nu} \, ,
\end{eqnarray}
where the  $\varepsilon_{\mu/\alpha}$ and $\varepsilon_{\mu\nu}$ are the polarization vectors.
In this article, we choose the components $\Pi_{+}(p^2)$ and $p^2\widetilde{\Pi}_{+}(p^2)$ to study the scalar, axialvector and tensor hidden-charm tetraquark states with the QCD sum rules.   In Table \ref{Current-Table}, we present the quark structures and corresponding interpolating currents for the hidden-charm tetraquark states.

Now we take a not very long  digression to discuss the feasibility of applying the QCD sum rules to study the tetraquark states and ${\bf 1}_c{\bf 1}_c$-type tetraquark states. We  usually perform Fierz rearrangement both in the color space  and Dirac-spinor  space to arrange  the diquark-antidiquark type currents  into a special superposition of the
  ${\bf 1}_c{\bf 1}_c$-type  currents \cite{Wang-tetra-formula}, for example,
\begin{eqnarray}\label{Fierz}
J^{SA}_{+,\mu}&=&\frac{1}{2\sqrt{2}}\Big\{\,i\bar{c}i\gamma_5 c\,\bar{d}\gamma_\mu u-i\bar{c} \gamma_\mu c\,\bar{d}i\gamma_5 u+\bar{c} u\,\bar{d}\gamma_\mu\gamma_5 c
-\bar{c} \gamma_\mu \gamma_5u\,\bar{d}c  \nonumber\\
&&  - i\bar{c}\gamma^\nu\gamma_5c\, \bar{d}\sigma_{\mu\nu}u+i\bar{c}\sigma_{\mu\nu}c\, \bar{d}\gamma^\nu\gamma_5u
- i \bar{c}\sigma_{\mu\nu}\gamma_5u\,\bar{d}\gamma^\nu c+i\bar{c}\gamma^\nu u\, \bar{d}\sigma_{\mu\nu}\gamma_5c   \,\Big\} \, .
\end{eqnarray}
In fact, the barrier or spatial separation between the diquark and antidiquark pair frustrates the Fierz rearrangements \cite{Wilczek-diquark,Polosa-diquark,Maiani-1903,Brodsky-PRL}.  If we neglect those frustrations, the diquark-antidiquark type currents and ${\bf 1}_c{\bf 1}_c$-type currents
can be rearranged into each other freely, they are all four-quark currents.

In the correlation functions for the ${\bf 1}_c{\bf 1}_c$-type currents,   Lucha, Melikhov and Sazdjian assert that  the Feynman diagrams can be divided into or separated into  factorizable  and nonfactorizable diagrams in the color space in the operator product expansion,
 the contributions  at the order $\mathcal{O}(\alpha_s^k)$ with $k\leq1$, which are factorizable in the color space, are exactly  canceled out    by the
 two-meson scattering states at the hadron side,
the nonfactorizable diagrams, if have a Landau singularity, begin to make contributions  to the (${\bf 1}_c{\bf 1}_c$-type) tetraquark  states,  the (${\bf 1}_c{\bf 1}_c$-type) tetraquark   states begin to receive contributions at the order $\mathcal{O}(\alpha_s^2)$ \cite{Chu-Sheng-PRD}.

In Ref.\cite{WangZG-Landau}, I refute the assertion of Lucha, Melikhov and Sazdjian in order and in details, and  use two examples to illustrate that
 the two-meson scattering states cannot saturate the QCD sum rules,  while
the ${\bf 1}_c{\bf 1}_c$-type tetraquark  states can saturate the QCD sum rules, the   ${\bf 1}_c{\bf 1}_c$-type tetraquark  states begin to receive contributions at the order $\mathcal{O}(\alpha_s^0/\alpha_s^1)$ rather than at the order $\mathcal{O}(\alpha_s^2)$.

The two-meson scattering state and ${\bf 1}_c{\bf 1}_c$-type tetraquark  state both have four valence quarks, which form two color-neutral clusters, we cannot distinguish the contributions  based on the two color-neutral clusters in the factorizable Feynman diagrams. It is questionable to assert that the factorizable Feynman diagrams only make contributions to the two-meson scattering states.  The Landau equation servers  as a kinematical equation in the momentum space, and is independent on the factorizable and nonfactorizable properties of the Feynman diagrams in the color space \cite{Landau}. The Landau equation cannot exclude the factorizable Feynman diagrams in the color space, as they are nonfactorizable in the momentum space and  also have Landau singularities.

 The quarks and gluons are confined objects, they cannot be put on the mass-shell, it is questionable  to  apply  the Landau equation to study the Feynman diagrams in the QCD sum rules. Furthermore,  we carry out the operator product expansion in the deep Euclidean region $P^2=-p^2\to \infty$, where the Landau singularities cannot exist.
There are other negative outcomes which involve applying the  Landau equation to study the (${\bf 1}_c{\bf 1}_c$-type) tetraquark  states, for detailed discussions about this subject, one can consult  Ref.\cite{WangZG-Landau}.

In Ref.\cite{Wang-Two-particle}, we study the $Z_c(3900)$ with the QCD sum rules in details by including all the two-particle scattering state contributions according to the Fierz rearrangement in Eq.\eqref{Fierz}, and observe that the two-particle scattering state contributions cannot saturate  the QCD sum rules at the hadron side,  the contribution of the $Z_c(3900)$ plays an un-substitutable role, we can saturate the QCD sum rules with or without the two-particle scattering state contributions.
We obtain the conclusion  that it is feasible to apply the QCD sum rules to study the diquark-antidiquark type  tetraquark states, which  begin to receive contributions at the order   $\mathcal{O}(\alpha_s^0)$, not at the order $\mathcal{O}(\alpha_s^2)$.

Now let us go back to correlation functions $\Pi(p)$, $\Pi_{\mu\nu}(p)$ and $\Pi_{\mu\nu\alpha\beta}(p)$ defined in Eq.\eqref{CF-Pi}.
At the QCD side, we carry out the operator product expansion for the correlation functions $\Pi(p)$, $\Pi_{\mu\nu}(p)$ and $\Pi_{\mu\nu\alpha\beta}(p)$ up to the vacuum condensates of dimension $10$ consistently, and take into account the vacuum condensates $\langle\bar{q}q\rangle$, $\langle\frac{\alpha_{s}GG}{\pi}\rangle$, $\langle\bar{q}g_{s}\sigma Gq\rangle$, $\langle\bar{q}q\rangle^2$, $g_s^2\langle\bar{q}q\rangle^2$,
$\langle\bar{q}q\rangle \langle\frac{\alpha_{s}GG}{\pi}\rangle$,  $\langle\bar{q}q\rangle  \langle\bar{q}g_{s}\sigma Gq\rangle$,
$\langle\bar{q}g_{s}\sigma Gq\rangle^2$ and $\langle\bar{q}q\rangle^2 \langle\frac{\alpha_{s}GG}{\pi}\rangle$, and obtain the QCD spectral densities $\rho(s)$ through dispersion relation. The contributions of the terms $g_s^2\langle\bar{q}q\rangle^2$ are tiny and neglected in most of the QCD sum rules, as they are not associated with the Borel parameters of the forms $\frac{1}{T^2}$, $\frac{1}{T^4}$, $\frac{1}{T^6}$, $\cdots$, which   amplify themselves at  small values of the $T^2$ and play   an important role in determining the Borel windows.   There are terms of the forms  $\langle\bar{q}_j\sigma_{\mu\nu}q_i \rangle$ and $\langle\bar{q}_j\gamma_{\mu}q_i\rangle$ in the full light quark propagators \cite{WangHuangtao-2014-PRD}, which  absorb the gluons  emitted from the other quark lines to form $\langle\bar{q}_j g_s G^a_{\alpha\beta} t^a_{mn}\sigma_{\mu\nu} q_i \rangle$ and $\langle\bar{q}_j\gamma_{\mu}q_ig_s D_\nu G^a_{\alpha\beta}t^a_{mn}\rangle$  to make contributions to the mixed condensate and four-quark condensate $\langle\bar{q}g_s\sigma G q\rangle$ and $g_s^2\langle\bar{q}q\rangle^2$, respectively.
 The four-quark condensate $g_s^2\langle \bar{q}q\rangle^2$ comes from the terms
$\langle \bar{q}\gamma_\mu t^a q g_s D_\eta G^a_{\lambda\tau}\rangle$, $\langle\bar{q}_jD^{\dagger}_{\mu}D^{\dagger}_{\nu}D^{\dagger}_{\alpha}q_i\rangle$  and
$\langle\bar{q}_jD_{\mu}D_{\nu}D_{\alpha}q_i\rangle$  rather than comes from the perturbative $\mathcal{O}(\alpha_s)$ corrections for the four-quark condensate $\langle \bar{q}q\rangle^2$, where $D_\alpha=\partial_\alpha-ig_sG^n_\alpha t^n$, $t^n=\frac{\lambda^n}{2}$,  the $\lambda^n$ is the Gell-Mann matrix. The strong coupling constant $\alpha_s(\mu)=\frac{g_s^2(\mu)}{4\pi}$ appears  at the tree level, which is energy scale dependent.  One can consult Ref.\cite{WangHuangtao-2014-PRD} for the technical details.  Furthermore, we  recalculate the higher  dimensional vacuum condensates using the formula $t^a_{ij}t^a_{mn}=-\frac{1}{6}\delta_{ij}\delta_{mn}+\frac{1}{2}\delta_{jm}\delta_{in}$, and obtain slightly  different expressions  compared to
  the old calculations.

As we have obtained both the hadron spectral representations and the QCD spectral representations, now we match  the hadron side with the QCD  side of the components $\Pi_{+}(p^2)$ and $p^2\widetilde{\Pi}_{+}(p^2)$ of the correlation functions $\Pi(p)$, $\Pi_{\mu\nu}(p)$ and $\Pi_{\mu\nu\alpha\beta}(p)$ below the continuum thresholds   $s_0$ and perform Borel transform  with respect to
 $P^2=-p^2$ to obtain  the  QCD sum rules:
\begin{eqnarray}\label{QCDSR}
\lambda^2_{Z^+}\, \exp\left(-\frac{M^2_{Z^+}}{T^2}\right)= \int_{4m_c^2}^{s_0} ds\, \rho(s) \, \exp\left(-\frac{s}{T^2}\right) \, .
\end{eqnarray}
 The explicit expressions of the QCD spectral densities $\rho(s)$ are available upon request by contacting me via  E-mail.

We derive Eq.\eqref{QCDSR} with respect to  $\tau=\frac{1}{T^2}$,  and obtain the QCD sum rules for
 the masses of the  scalar,  axialvector  and tensor hidden-charm tetraquark states $Z_c$ or $X$,
 \begin{eqnarray}
 M^2_{Z^+}&=& -\frac{\int_{4m_c^2}^{s_0} ds\frac{d}{d \tau}\rho(s)\exp\left(-\tau s \right)}{\int_{4m_c^2}^{s_0} ds \rho(s)\exp\left(-\tau s\right)}\, .
\end{eqnarray}

\section{Numerical results and discussions}
In this article, we neglect the small $u$ and $d$ quark masses. The  heavy quark ($\overline{MS}$) mass $m_c(\mu)$ and the vacuum condensates depend on the energy scale $\mu$, so the QCD spectral densities $\rho(s,\mu)$ depend on the energy scale $\mu$, the thresholds $4m_c^2(\mu)$ also depend on the energy scale $\mu$, we cannot extract the masses of the hidden-charm tetraquark states  from the energy scale independent QCD sum rules, and have to choose the best  energy scales to extract the tetraquark masses.
The energy-scale dependence of  the input parameters at the QCD side can be written as
\begin{eqnarray}
\langle\bar{q}q \rangle(\mu)&=&\langle\bar{q}q \rangle({\rm 1GeV})\left[\frac{\alpha_{s}({\rm 1GeV})}{\alpha_{s}(\mu)}\right]^{\frac{12}{33-2n_f}}\, , \nonumber\\
 \langle\bar{q}g_s \sigma Gq \rangle(\mu)&=&\langle\bar{q}g_s \sigma Gq \rangle({\rm 1GeV})\left[\frac{\alpha_{s}({\rm 1GeV})}{\alpha_{s}(\mu)}\right]^{\frac{2}{33-2n_f}}\, , \nonumber\\
 m_c(\mu)&=&m_c(m_c)\left[\frac{\alpha_{s}(\mu)}{\alpha_{s}(m_c)}\right]^{\frac{12}{33-2n_f}} \, ,\nonumber\\
\alpha_s(\mu)&=&\frac{1}{b_0t}\left[1-\frac{b_1}{b_0^2}\frac{\log t}{t} +\frac{b_1^2(\log^2{t}-\log{t}-1)+b_0b_2}{b_0^4t^2}\right]\, ,
\end{eqnarray}
 from the renormalization group equation,  where $t=\log \frac{\mu^2}{\Lambda_{QCD}^2}$, $b_0=\frac{33-2n_f}{12\pi}$, $b_1=\frac{153-19n_f}{24\pi^2}$, $b_2=\frac{2857-\frac{5033}{9}n_f+\frac{325}{27}n_f^2}{128\pi^3}$,  $\Lambda_{QCD}=210\,\rm{MeV}$, $292\,\rm{MeV}$  and  $332\,\rm{MeV}$ for the flavors  $n_f=5$, $4$ and $3$, respectively  \cite{PDG,Narison-mix}. In the present work, as the $c$-quark is involved, we take the flavor $n_f=4$.

 At the initial points, we take  the standard values of the vacuum condensates $\langle
\bar{q}q \rangle=-(0.24\pm 0.01\, \rm{GeV})^3$,   $\langle
\bar{q}g_s\sigma G q \rangle=m_0^2\langle \bar{q}q \rangle$,
$m_0^2=(0.8 \pm 0.1)\,\rm{GeV}^2$,  $\langle \frac{\alpha_s
GG}{\pi}\rangle=(0.33\,\rm{GeV})^4 $    at the energy scale  $\mu=1\, \rm{GeV}$
\cite{SVZ79,Reinders85,Colangelo-Review}, and take the $\overline{MS}$ mass $m_{c}(m_c)=(1.275\pm0.025)\,\rm{GeV}$ from the Particle Data Group \cite{PDG}.

The hidden-bottom or hidden-charm tetraquark states can be described by  a double-well potential model,  the heavy  quark $Q$ (heavy antiquark $\overline{Q}$) serves as a static well potential and attracts  the light quark $q$ (light antiquark $\bar{q}$)  to form a diquark (antidiquark) in the color antitriplet (triplet) channel.
The diquark-antidiquark type tetraquark states, which are excellent candidates for the $X$, $Y$, $Z$ states,  are characterized by  the effective heavy quark mass ${\mathbb{M}}_Q$ and the virtuality $V=\sqrt{M^2_{X/Y/Z}-(2{\mathbb{M}}_Q)^2}$. If we choose the energy  scale \cite{Wang-tetra-formula,WangHuang-2014-NPA},
 \begin{eqnarray}
 \mu^2&=&V^2={\mathcal{O}}(T^2)\, ,
 \end{eqnarray}
then we obtain the formula $\mu=\sqrt{M^2_{X/Y/Z}-(2{\mathbb{M}}_Q)^2}$ to determine the best  energy scales for  the QCD sum rules. At the ideal energy scales, we can enhance the pole contributions at the hadron side remarkably and improve the convergent behaviors of the operator product expansion at the QCD side remarkably.
In the present work, we use the energy scale formula $\mu=\sqrt{M^2_{X/Y/Z}-(2{\mathbb{M}}_c)^2}$ with the updated effective $c$-quark mass ${\mathbb{M}}_c=1.82\,\rm{GeV}$ to determine the ideal energy scales for the QCD sum rules \cite{WangZG-eff-Mc}.

At the QCD side, we carry out the operator product expansion at the large space-like region,
\begin{eqnarray}
x_0\sim |\vec{x}|\sim \frac{1}{\sqrt{P^2}}\ll \frac{1}{\Lambda_{QCD}}\, ,
\end{eqnarray}
the anomalous dimensions $\gamma_J$ of the interpolating currents $J(x)$, $J_\mu(x)$ and $J_{\mu\nu}(x)$ lead to a factor multiplying the correlation functions at the QCD side \cite{SB-Ioffe}. After the Borel transformation, the factor is changed to
\begin{eqnarray}
\left[\frac{\alpha_s(P^2)}{\alpha_s(\mu^2)} \right]^{2\gamma_J} &\to& \left[\frac{\alpha_s(T^2)}{\alpha_s(\mu^2)} \right]^{2\gamma_J}\, .
\end{eqnarray}
If we choose $\mu^2={\mathcal{O}}(T^2)$, the factor is neglectful.
In fact, the present approach weakens  the energy scale dependence of the QCD sum rules, although we cannot obtain energy scale independent QCD sum rules.

The continuum threshold parameters are not completely free parameters, and cannot be determined by the QCD sum rules themselves completely. We often consult the experimental data in choosing the continuum threshold parameters.
The  $Z_c(4430)$ can be assigned to be the first radial excitation of the $Z_c(3900)$ according to the
analogous decays,
\begin{eqnarray}
Z_c^\pm(3900)&\to&J/\psi\pi^\pm\, , \nonumber \\
Z_c^\pm(4430)&\to&\psi^\prime\pi^\pm\, ,
\end{eqnarray}
and the  analogous mass gaps  $M_{Z_c(4430)}-M_{Z_c(3900)}=591\,\rm{MeV}$ and $M_{\psi^\prime}-M_{J/\psi}=589\,\rm{MeV}$ from the Particle Data Group \cite{PDG,Maiani-Z4430-1405,Nielsen-1401,WangZG-Z4430-CTP}. Thereafter, we will use the superscripts $\pm$ to denote the electric charges.
The energy gap $M_{X(4500)}-M_{X(3915)}=588\,\rm{MeV}$ from the Particle Data Group \cite{PDG}, the $X(3915)$ and $X(4500)$ can be assigned as  the ground state and the first radial excited state of the axialvector-diquark-axialvector-antidiquark type scalar $cs\bar{c}\bar{s}$ tetraquark states \cite{Lebed-X3915}, the QCD sum rules also support such assignments \cite{WangZG-X4500}.
Recently, the LHCb collaboration performed an angular analysis of the $B^0\to J/\psi K^+\pi^-$ decays, and observed two possible  structures near $m(J/\psi \pi^-)=4200 \,\rm{MeV}$ and $4600\,\rm{MeV}$, respectively \cite{LHCb-Z4600}.
  There have been two tentative assignments for the  structure $Z_c(4600)$, the  vector tetraquark state with $J^{PC}=1^{--}$ \cite{Wang-Z4600-V} and the first radial excited  axialvector tetraquark state with $J^{PC}=1^{+-}$  \cite{ChenHX-Z4600-A,WangZG-axial-Z4600}. If the dominant Fock component of the $Z_c(4600)$ is an axialvector tetraquark state,  the energy gap between the ground state $Z_c(4020)$ and the first radial excited state $Z_c(4600)$ is about $M_{Z_c(4600)}-M_{Z_c(4020)}=576\,\rm{MeV}$ from the Particle Data Group \cite{PDG}.

In the present work, we tentatively choose the continuum threshold parameters as $\sqrt{s_0}=M_{Z}+0.58/0.59\,\rm{GeV}$ and vary the continuum threshold parameters $s_0$ and Borel parameters $T^2$ to satisfy
the following four   criteria:\\
$\bullet$ Pole dominance at the hadron  side;\\
$\bullet$ Convergence of the operator product expansion;\\
$\bullet$ Appearance of the Borel platforms;\\
$\bullet$ Satisfying the  energy scale formula,\\
  via trial  and error.

The pole dominance at the hadron  side and convergence of the operator product expansion at the QCD side are two basic  criteria for the QCD sum rules, we should satisfy the two basic  criteria to obtain reliable QCD sum rules. Furthermore, we should obtain Borel platforms to avoid additional uncertainties originate  from the Borel parameters.
The pole contributions (PC) or ground state tetraquark contributions are defined by
\begin{eqnarray}
{\rm{PC}}&=&\frac{\int_{4m_{c}^{2}}^{s_{0}}ds\rho\left(s\right)\exp\left(-\frac{s}{T^{2}}\right)} {\int_{4m_{c}^{2}}^{\infty}ds\rho\left(s\right)\exp\left(-\frac{s}{T^{2}}\right)}\, ,
\end{eqnarray}
 while the contributions of the vacuum condensates $D(n)$ of dimension $n$ are defined by
\begin{eqnarray}
D(n)&=&\frac{\int_{4m_{c}^{2}}^{s_{0}}ds\rho_{n}(s)\exp\left(-\frac{s}{T^{2}}\right)}
{\int_{4m_{c}^{2}}^{s_{0}}ds\rho\left(s\right)\exp\left(-\frac{s}{T^{2}}\right)}\, ,
\end{eqnarray}
as we only study the ground state contributions. At the QCD side of the correlation functions $\Pi(p)$, $\Pi_{\mu\nu}(p)$ and $\Pi_{\mu\nu\alpha\beta}(p)$, there are four full quark propagators, i.e.  two heavy  quark propagators and two light quark propagators.  If each heavy quark line emits a gluon and each light quark line contributes  a quark-antiquark  pair, we obtain a quark-gluon  operator $GG\bar{u}u \bar{d}d$   of dimension 10, so we should calculate  the vacuum condensates at least
up to dimension 10 to judge the convergent behavior of the operator product expansion. The vacuum condensates  $\langle\bar{q}g_{s}\sigma Gq\rangle^2$ and $\langle\bar{q}q\rangle^2 \langle\frac{\alpha_{s}GG}{\pi}\rangle$ are associated with $\frac{1}{T^2}$, $\frac{1}{T^4}$ and  $\frac{1}{T^6}$, they play an important role in determining the Borel windows, although in the Borel windows, they are of minor importance. In the present work, we require  the contributions $|D(10)|\sim 1\%$ at the Borel windows.

Finally, we obtain the Borel windows, continuum threshold parameters, energy scales of the QCD spectral densities,  pole contributions, and the contributions of the vacuum condensates of dimension $10$, which are shown explicitly in Table \ref{BorelP}.
From the Table,  we can see that the pole contributions are about $(40-60)\%$ at the hadron side, the pole dominance condition  is well satisfied.
At the QCD side, the contributions of the vacuum condensates of dimension $10$ are $|D(10)|\leq 1 \%$ or $\ll 1\%$ except for $|D(10)|< 2 \%$ for the $[uc]_{\widetilde{V}}[\overline{dc}]_{V}-[uc]_{V}[\overline{dc}]_{\widetilde{V}}$ tetraquark state with the spin-parity-charge-conjugation $J^{PC}=1^{++}$, the convergent behavior of the operator product  expansion is very good.

\begin{figure}
 \centering
 \includegraphics[totalheight=6cm,width=7cm]{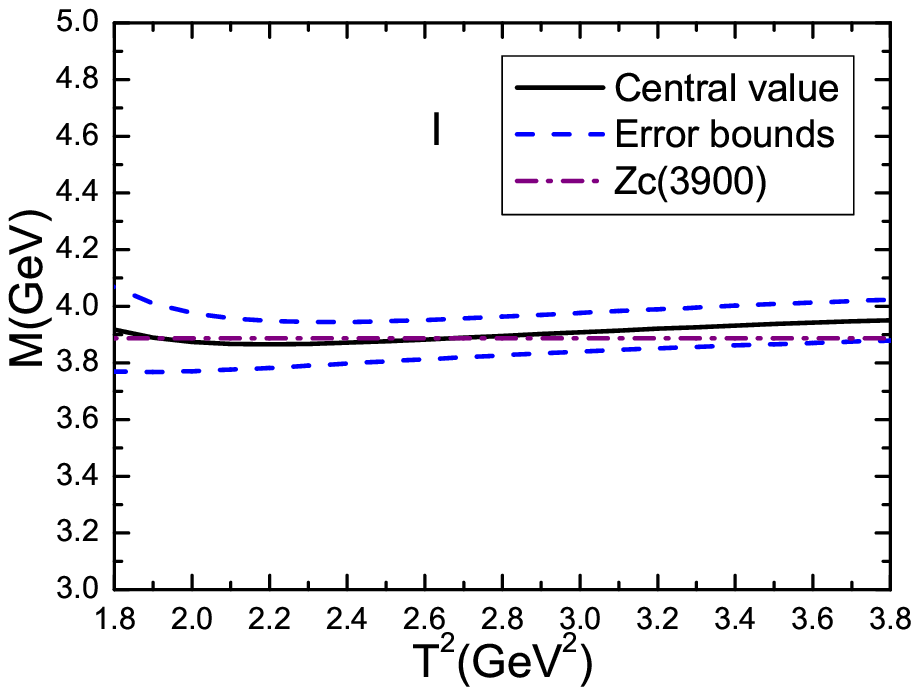}
 \includegraphics[totalheight=6cm,width=7cm]{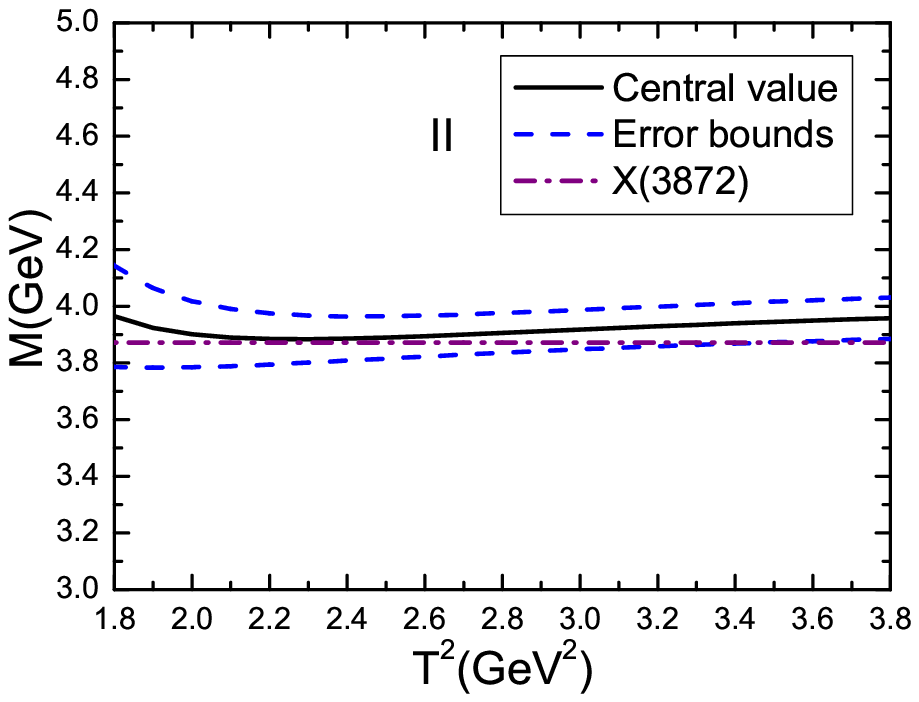}
 \caption{ The masses of the  $[uc]_S[\overline{dc}]_{A}-[uc]_{A}[\overline{dc}]_S$(I) and $[uc]_S[\overline{dc}]_{A}+[uc]_{A}[\overline{dc}]_S$(II) axialvector tetraquark states   with variations  of the Borel parameters $T^2$.  }\label{mass-1-fig}
\end{figure}

We take into account all the uncertainties of the input parameters and obtain the masses and pole residues of the scalar, axialvector, tensor hidden-charm  tetraquark states, which are shown explicitly in Table \ref{mass-Table}. From  Tables \ref{BorelP}--\ref{mass-Table}, we can see that the energy scale formula $\mu=\sqrt{M^2_{X/Y/Z}-(2{\mathbb{M}}_c)^2}$ is well satisfied.
 In  Fig.\ref{mass-1-fig}, we plot the masses of the  $[uc]_S[\overline{dc}]_{A}-[uc]_{A}[\overline{dc}]_S$ and $[uc]_S[\overline{dc}]_{A}+[uc]_{A}[\overline{dc}]_S$ axialvector tetraquark states with variations of the Borel parameters at much larger ranges than the Borel widows as an example. From the figure, we can see that there appear platforms in the Borel windows indeed.       Now the four criteria of the QCD sum rules are all satisfied, and we expect to make reliable predictions.

In Table \ref{Assignments-Table}, we present the possible assignments of the ground state hidden-charm tetraquark states.
There are  one scalar tetraquark candidate with the spin-parity-charge-conjugation $J^{PC}=0^{++}$ for the $X(3860)$,
one scalar tetraquark candidate with the   $J^{PC}=0^{++}$ for the $X(3915)$,
one axialvector tetraquark candidate with the  $J^{PC}=1^{++}$ for the $X(3872)$,
one axialvector tetraquark candidate with the  $J^{PC}=1^{+-}$ for the $Z_c(3900)$,
one axialvector tetraquark candidate with the  $J^{PC}=1^{+-}$ for the $Z_c(4600)$,
three axialvector tetraquark candidates with the  $J^{PC}=1^{+-}$ for the $Z_c(4020)$ and $Z_c(4055)$,
two axialvector tetraquark candidates with the  $J^{PC}=1^{++}$ for the $Z_c(4050)$,
one tensor  tetraquark candidate with the  $J^{PC}=2^{++}$ for the $Z_c(4050)$.

In 2008, the Belle collaboration  observed two resonance-like structures, which are known as the $Z_c(4050)$ and $Z_c(4250)$ now,  in the $\pi^+\chi_{c1}$ mass spectrum  in the decays $\bar{B}^0\to K^- \pi^+ \chi_{c1}$ with a statistical significance  exceeds $5 \sigma$ \cite{Belle-chipi}.
In 2014, the Belle collaboration observed an evidence for the $Z_c(4055)$ in the $\pi^\pm\psi^\prime$ mass spectrum in the decays $Y(4360)\to \pi^+\pi^-\psi^\prime$
 with a statistical significance of  $3.5\sigma$ \cite{Belle-Z4055}. The $Z^+_c(4050)$, $Z_c^\pm(4055)$ and $Z_c^+(4250)$ have not been confirmed by other experiments yet.

 The $Z^+_c(4050)$, $Z_c^\pm(4055)$ and $Z_c^+(4250)$ are not necessary  to have the definite charge conjugation $C=+$, $-$ and $+$, respectively, as they are not the charge conjugation eigenstates. The possible quantum numbers of the $Z^+_c(4050)$ and $Z_c^+(4250)$ are the spin-parity  $J^{P}=0^+$, $1^-$, $1^+$ and $2^+$. From  Table \ref{Assignments-Table}, we can see  that there is no candidate for the $Z_c^+(4250)$ in the case of the spin-parity $J^{P}=0^+$,  $1^+$ and $2^+$. In Ref.\cite{WangZG-V-tetra}, we observe that the $Z_c(4250)$ can be assigned to be a vector tetraquark state with a relative P-wave between the diquark and antidiquark constituents based  on the predictions of the QCD sum rules.

From Table \ref{Assignments-Table}, we can see that the lowest tetraquark state with the $J^{PC}=1^{+-}$ has a mass about $3.9\,\rm{GeV}$, if the energy gap between the ground state and the first radial excited state is about $0.5-0.6\,\rm{GeV}$, the first radial excitation of the axialvector tetraquark state has a mass about $4.4-4.5\,\rm{GeV}$, which happens to  coincide  with the experimental mass of the $Z_c(4430)$, see Table 1. The $Z_c(3900)$ and $Z_c(4430)$ can be assigned to be the ground state and the first radial excited axialvector tetraquark states with the $J^{PC}=1^{+-}$, respectively \cite{Maiani-Z4430-1405,Nielsen-1401,WangZG-Z4430-CTP}, the QCD sum rules supports such assignments \cite{WangZG-Z4430-CTP}. The QCD sum rules also support assigning the $Z_c(4600)$ to be the first radial excited state of the $Z_c(4020)$ \cite{ChenHX-Z4600-A,WangZG-axial-Z4600}.

The $Z_c(4200)$ has the spin-parity-charge-conjugation $J^{PC}=1^{+-}$, there is no room to accommodate it in the scenario of pure tetraquark  state. The pure axialvector tetraquark states have the masses about $3.9\,\rm{GeV}$, $4.0\,\rm{GeV}$, $4.7\,\rm{GeV}$ and $5.5\,\rm{GeV}$, respectively,  which are inconsistent with  the experimental value $M_{Z_c(4200)} = 4196^{+31}_{-29}{}^{+17}_{-13} \,\rm{MeV}$ \cite{PDG}. However, a mixing $[uc]_S[\overline{dc}]_{A}-[uc]_{A}[\overline{dc}]_S$ or
$[uc]_{A}[\overline{dc}]_{A}$  or
$[uc]_S[\overline{dc}]_{\widetilde{A}}-[uc]_{\widetilde{A}}[\overline{dc}]_S$ or
$[uc]_{\widetilde{A}}[\overline{dc}]_{A}-[uc]_{A}[\overline{dc}]_{\widetilde{A}}$ $\oplus$
$[uc]_{\widetilde{V}}[\overline{dc}]_{V}+[uc]_{V}[\overline{dc}]_{\widetilde{V}}$ 
 or $[uc]_{V}[\overline{dc}]_{V}$ or
$[uc]_P[\overline{dc}]_{V}+[uc]_{V}[\overline{dc}]_P$  type axialvector tetraquark state with the spin-parity-charge-conjugation  $J^{PC}=1^{+-}$ may be have a mass about $4.2\,\rm{GeV}$.  On the other hand, in Ref.\cite{WangZG-Z4200-IJMPA}, we observe that the $Z_c(4200)$ can be assigned to be the color-octet-color-octet ($ {\bf8}_c{\bf 8}_c$) type tetraquark  state by calculating the mass and width with the QCD sum rules.

If the $Z_c(4100)$ has the   spin-parity-charge-conjugation  $J^{PC}=0^{++}$, it cannot be a  pure scalar tetraquark candidate according to the predictions presented in Table \ref{Assignments-Table}. In Ref.\cite{WangZG-Z4100-1903}, we observe that  if we introduce the mixing effects,  a mixing $[uc]_{\tilde{A}}[\overline{dc}]_{\tilde{A}}\oplus[uc]_{\tilde{V}}[\overline{dc}]_{\tilde{V}}$ type scalar tetraquark state can  have a mass about $4.1\,\rm{GeV}$.
From Table  \ref{Assignments-Table}, we can see that  a mixing $[uc]_{S}[\overline{dc}]_{S}$ or  $[uc]_{A}[\overline{dc}]_{A}$ or $[uc]_{\tilde{A}}[\overline{dc}]_{\tilde{A}}$ $\oplus$  $[uc]_{V}[\overline{dc}]_{V}$ or $[uc]_{\tilde{V}}[\overline{dc}]_{\tilde{V}}$ or
$[uc]_{P}[\overline{dc}]_{P}$ type scalar tetraquark state with  suitable mixing angles can have  a mass about $4.1\,\rm{GeV}$.

On the other hand, in Refs.\cite{WangZG-V-tetra,WangZG-V-Y4660}, we observe that the lowest vector tetraquark state from the QCD sum rules has a mass about $4.24\pm0.10\,\rm{GeV}$, which lies above the $Z_c(4100)$, the $Z_c(4100)$ is unlikely to be a vector tetraquark state.

If we abandon the constraint $\mu=\sqrt{M^2_{X/Y/Z}-(2{\mathbb{M}}_c)^2}$ in choosing the suitable  energy scales of the QCD spectral densities,   and assign the $Z_c(4200)$ to be the $[uc]_S[\overline{dc}]_{A}-[uc]_{A}[\overline{dc}]_S$  type  hidden-charm tetraquark state with the quantum numbers  $J^P=1^{+-}$, we can reproduce
the experimental value $M_{Z_c(4200)} = 4196^{+31}_{-29}{}^{+17}_{-13} \,\rm{MeV}$  from the Belle collaboration by choosing the energy scale $\mu=1.2\,\rm{GeV}$ \cite{WangZG-Z4200-IJMPA}.
Furthermore, if we  assign the $Z_c(4100)$ to be the $[uc]_{\tilde{A}}[\overline{dc}]_{\tilde{A}}$ type hidden-charm tetraquark state with the quantum numbers  $J^P=0^{++}$,
we obtain the prediction $M_{Z_c(4100)}=4.12\pm0.08\,\rm{GeV}$ by choosing the energy scale $\mu=1.4\,\rm{GeV}$ \cite{WangZG-Z4100-1903}, which is in excellent agreement with the experimental value $M_{Z_c}=4096 \pm 20^{+18}_{-22}\,\rm{MeV}$ from the LHCb  collaboration \cite{LHCb-Z4100}. However, the energy scales $\mu=1.2\,\rm{GeV}$  and $1.4\,\rm{GeV}$ are chosen by hand, and are inconsistent with  the energy scales presented in Table \ref{BorelP}, furthermore,  the two energy scales are even
inconsistent with each other.
Without introducing mixing effects among the tetraquark states, the QCD sum rules disfavor  assigning the $Z_c(4100)$ and $Z_c(4200)$
to be the hidden-charm tetraquark states.

The $X(3940)$ and $X(4160)$ were observed in the processes $e^+e^-\to J/\psi D^*\bar{D}$ and $J/\psi D^*\bar{D}^*$ respectively by the Belle collaboration \cite{Belle-X3940,Belle-X4160}.
The absence of signals for any of the known non-zero
spin charmonium states in the distribution of masses recoiling from the $J/\psi$ in the inclusive spectrum
 provides circumstantial evidence for zero spin assignments
for the $X(3940)$ and $X(4160)$. If they have zero spin, the observed decays $X(3940)\to D^*\bar{D}$ and $X(4160) \to D^*\bar{D}^*$
 support that they have the quantum numbers $J^{PC}=0^{-+}$, however, the assignment of the  $J^{PC}=0^{++}$ for the $X(4160)$ cannot be excluded. The calculations based on the QCD sum rules indicate that the tetraquark states with the $J^{PC}=0^{-+}$ have much larger masses than the $X(4160)$, the numerical results will be presented elsewhere. On the other hand, if the $X(4160)$ is an scalar hidden-charm tetraquark state, we have to introduce the mixing effects to account for the mass. In summary, there are no rooms  to accommodate the $X(3940)$ and $X(4160)$ in the scenario of tetraquark  states without fine tuning.
 They may be  the conventional $\eta_c(\rm 3S)$ and $\eta_c(\rm 4S)$ with the quantum numbers $J^{PC}=0^{-+}$, respectively  \cite{Rosner-X3940-KTChao-X4160}. However, the measured masses lie far below expectations based on the potential models. More theoretical and experimental works are still  needed to explore the nature of the $X(3940)$ and $X(4160)$.

The hidden-charm tetraquark masses obtained in the present work can be confronted to the experimental data  at the BESIII, LHCb, Belle II,  CEPC, FCC, ILC
 in the future, and shed light on the nature of the exotic $X$, $Y$, $Z$ particles.

Although the mass is the basic and most important parameter in describing a hadron, we cannot assign a hadron unambiguously based on the mass alone, we need other quantum numbers, such as the spin, parity  and conjugation, and have to study its productions and decays. For example, the $X(3872)$ can be assigned to be the tetraquark state or ${\bf1}_c{\bf 1}_c$-type tetraquark state based on the predicted mass from the QCD sum rules, however, without introducing the $c\bar{c}$ or $\chi_{c1}^{\prime}$ component, we cannot describe   its productions  at the hadron colliders and the hadronic decays $X(3872)\to J/\psi \pi^+\pi^-$ and $J/\psi\pi^+\pi^-\pi^0$ \cite{KTChao-3872,MNielsen-cc-3872}.

 We can take the pole residues $\lambda_Z$ as the basic input parameters to study the two-body
 strong decays of those hidden-charm tetraquark states,
 \begin{eqnarray}
Z_c^\pm(1^{+-}) &\to&\pi^{\pm}J/\psi\,  ,  \,\pi^{\pm}\psi^\prime\,  ,  \, \pi^{\pm} h_c \, ,  \, \rho^{\pm} \eta_c  \, , \, (D\bar{D}^{*})^\pm\, ,\, (D^{*}\bar{D})^\pm\, ,\, (D^{*}\bar{D}^{*})^\pm\, , \nonumber\\
Z_c^\pm(0^{++}) &\to&\pi^{\pm}\eta_c\,  ,  \, \pi^{\pm} \chi_{c1} \, ,  \, \rho^{\pm} J/\psi \, , \, \rho^{\pm} \psi^\prime \, , \,(D\bar{D})^\pm\, ,\, (D^{*}\bar{D}^{*})^\pm\, , \nonumber\\
Z_c^\pm(1^{++}) &\to& \pi^{\pm} \chi_{c1} \, ,  \, \rho^{\pm} J/\psi \, , \, \rho^{\pm} \psi^\prime \, , \,(D\bar{D}^{*})^\pm\, ,\, (D^{*}\bar{D})^\pm\, ,\, (D^{*}\bar{D}^{*})^\pm\, , \nonumber\\
Z_c^\pm(2^{++}) &\to&\pi^{\pm}\eta_c \,  ,  \, \pi^{\pm} \chi_{c1}  \, ,  \, \rho^{\pm} J/\psi \, , \, \rho^{\pm} \psi^\prime \, , \,(D\bar{D})^\pm\, ,\, (D^{*}\bar{D}^{*})^\pm\, ,
\end{eqnarray}
 \begin{eqnarray}
Z_c^0(1^{+-}) &\to&\pi^{0}J/\psi\,  ,  \,\pi^{0}\psi^\prime\,  ,  \, \pi^{0} h_c \, ,  \, \rho^{0} \eta_c  \, , \, (D\bar{D}^{*})^0\, ,\, (D^{*}\bar{D})^0\, ,\, (D^{*}\bar{D}^{*})^0\, , \nonumber\\
Z_c^0(0^{++}) &\to&\pi^{0}\eta_c\,  ,  \, \pi^{0} \chi_{c1} \, ,  \, \rho^{0} J/\psi \, , \, \rho^{0} \psi^\prime \, , \,(D\bar{D})^0\, ,\, (D^{*}\bar{D}^{*})^0\, , \nonumber\\
Z_c^0(1^{++}) &\to& \pi^{0} \chi_{c1} \, ,  \, \rho^{0} J/\psi \, , \, \rho^{0} \psi^\prime \, , \,(D\bar{D}^{*})^0\, ,\, (D^{*}\bar{D})^0\, ,\, (D^{*}\bar{D}^{*})^0\, , \nonumber\\
Z_c^0(2^{++}) &\to&\pi^{0}\eta_c \,  ,  \, \pi^{0} \chi_{c1}  \, ,  \, \rho^{0} J/\psi \, , \, \rho^{0} \psi^\prime \, , \,(D\bar{D})^0\, ,\, (D^{*}\bar{D}^{*})^0\, ,
\end{eqnarray}
\begin{eqnarray}
X(1^{+-}) &\to&\eta J/\psi\,  ,  \,\eta\psi^\prime\,  ,  \,\eta h_c \, ,  \, \omega \eta_c  \, , \, (D\bar{D}^{*})^0\, ,\, (D^{*}\bar{D})^0\, ,\, (D^{*}\bar{D}^{*})^0\, , \nonumber\\
X(0^{++}) &\to&\eta\eta_c\,  ,  \, \eta \chi_{c1} \, ,  \, \omega J/\psi \, , \, \omega\psi^\prime \, , \,(D\bar{D})^0\, ,\, (D^{*}\bar{D}^{*})^0\, , \nonumber\\
X(1^{++}) &\to& \eta \chi_{c1} \, ,  \, \omega J/\psi \, , \, \omega \psi^\prime \, , \,(D\bar{D}^{*})^0\, ,\, (D^{*}\bar{D})^0\, ,\, (D^{*}\bar{D}^{*})^0\, , \nonumber\\
X(2^{++}) &\to&\eta\eta_c \,  ,  \, \eta \chi_{c1}  \, ,  \, \omega J/\psi \, , \, \omega \psi^\prime \, , \,(D\bar{D})^0\, ,\, (D^{*}\bar{D}^{*})^0\, ,
\end{eqnarray}
with the three-point QCD sum rules or the light-cone QCD sum rules, and obtain the partial decay widths  to diagnose the nature of the $Z_c$ and $X$ states.
For example,   we tentatively assign the $Z_c^+(3900)$ to be  the $[uc]_S[\overline{dc}]_{A}-[uc]_{A}[\overline{dc}]_S$  type axialvector  tetraquark  state,   study  the two-body strong decays $Z_c^+(3900)\to J/\psi\pi^+$, $\eta_c\rho^+$, $D^+ \bar{D}^{*0}$, $\bar{D}^0 D^{*+}$ with the QCD sum rules  based on solid quark-hadron duality, and produce the experimental value of  the total width \cite{WangZhang-Solid}. Experimentally, the BESIII collaboration measured the ratios of the partial widths of the decays $Z_c(3900/4020) \to \eta_c\rho\, , \,J/\psi \pi$    at the $90\%$ C.L. \cite{CZYuan-Decay}.

\begin{table}
\begin{center}
\begin{tabular}{|c|c|c|c|c|c|c|c|c|}\hline\hline
 $Z_c$($X_c$)                                         & $J^{PC}$ & $T^2 (\rm{GeV}^2)$ & $\sqrt{s_0}(\rm GeV) $      &$\mu(\rm{GeV})$   &pole         &$|D(10)|$ \\ \hline

$[uc]_{S}[\overline{dc}]_{S}$                         & $0^{++}$ & $2.7-3.1$          & $4.40\pm0.10$               &$1.3$             &$(40-63)\%$  &$<1\%$   \\

$[uc]_{A}[\overline{dc}]_{A}$                         & $0^{++}$ & $2.8-3.2$          & $4.52\pm0.10$               &$1.5$             &$(40-63)\%$  &$\leq1\%$    \\

$[uc]_{\tilde{A}}[\overline{dc}]_{\tilde{A}}$         & $0^{++}$ & $3.1-3.5$          & $4.55\pm0.10$               &$1.6$             &$(42-62)\%$  &$<1\%$    \\

$[uc]_{V}[\overline{dc}]_{V}$                         & $0^{++}$ & $3.7-4.1$          & $5.22\pm0.10$               &$2.9$             &$(41-60)\%$  &$\ll1\%$    \\

$[uc]_{\tilde{V}}[\overline{dc}]_{\tilde{V}}$         & $0^{++}$ & $4.9-5.7$          & $5.90\pm0.10$               &$3.9$             &$(41-61)\%$  &$\ll 1\%$   \\

$[uc]_{P}[\overline{dc}]_{P}$                         & $0^{++}$ & $5.2-6.0$          & $6.03\pm0.10$               &$4.1$             &$(40-60)\%$  &$\ll1\%$    \\ \hline

$[uc]_S[\overline{dc}]_{A}-[uc]_{A}[\overline{dc}]_S$ & $1^{+-}$ & $2.7-3.1$          & $4.40\pm0.10$               &$1.4$             &$(40-63)\%$  &$<1\%$    \\

$[uc]_{A}[\overline{dc}]_{A}$                         & $1^{+-}$ & $3.3-3.7$          & $4.60\pm0.10$               &$1.7$             &$(40-59)\%$  &$\ll 1\%$  \\

$[uc]_S[\overline{dc}]_{\widetilde{A}}-[uc]_{\widetilde{A}}[\overline{dc}]_S$     & $1^{+-}$ & $3.3-3.7$     & $4.60\pm0.10$     &$1.7$      &$(40-59)\%$ &$\ll 1\%$  \\

$[uc]_{\widetilde{A}}[\overline{dc}]_{A}-[uc]_{A}[\overline{dc}]_{\widetilde{A}}$ & $1^{+-}$ & $3.2-3.6$     & $4.60\pm0.10$     &$1.7$      &$(41-61)\%$ &$\ll 1\%$ \\

$[uc]_{\widetilde{V}}[\overline{dc}]_{V}+[uc]_{V}[\overline{dc}]_{\widetilde{V}}$ & $1^{+-}$ & $3.7-4.1$     & $5.25\pm0.10$     &$2.9$      &$(41-60)\%$ &$\ll 1\%$ \\

$[uc]_{V}[\overline{dc}]_{V}$                         & $1^{+-}$ & $5.1-5.9$          & $6.00\pm0.10$               &$4.1$             &$(41-60)\%$  &$\ll 1\%$  \\

$[uc]_P[\overline{dc}]_{V}+[uc]_{V}[\overline{dc}]_P$ & $1^{+-}$ & $5.1-5.9$          & $6.00\pm0.10$               &$4.1$             &$(41-60)\%$  &$\ll1\%$    \\
\hline

$[uc]_S[\overline{dc}]_{A}+[uc]_{A}[\overline{dc}]_S$ & $1^{++}$ & $2.7-3.1$          & $4.40\pm0.10$               &$1.4$             &$(40-62)\%$  &$\ll 1\%$  \\

$[uc]_S[\overline{dc}]_{\widetilde{A}}+[uc]_{\widetilde{A}}[\overline{dc}]_S$     & $1^{++}$ & $3.3-3.7$     & $4.60\pm0.10$     &$1.7$      &$(40-59)\%$ &$\ll 1\%$  \\

$[uc]_{\widetilde{V}}[\overline{dc}]_{V}-[uc]_{V}[\overline{dc}]_{\widetilde{V}}$ & $1^{++}$ & $2.8-3.2$     & $4.62\pm0.10$     &$1.8$      &$(40-63)\%$ &$<2\%$ \\

$[uc]_{\widetilde{A}}[\overline{dc}]_{A}+[uc]_{A}[\overline{dc}]_{\widetilde{A}}$ & $1^{++}$ & $4.6-5.3$     & $5.73\pm0.10$     &$3.7$      &$(40-60)\%$ &$\ll 1\%$ \\

$[uc]_P[\overline{dc}]_{V}-[uc]_{V}[\overline{dc}]_P$ & $1^{++}$ & $5.1-5.9$          & $6.00\pm0.10$               &$4.1$             &$(40-60)\%$  &$\ll1\%$    \\
\hline

$[uc]_{A}[\overline{dc}]_{A}$                         & $2^{++}$ & $3.3-3.7$          & $4.65\pm0.10$               &$1.8$             &$(40-60)\%$       &$<1\%$ \\

$[uc]_{V}[\overline{dc}]_{V}$                         & $2^{++}$ & $5.0-5.8$          & $5.95\pm0.10$               &$4.0$             &$(40-60)\%$       &$\ll1\%$ \\
\hline\hline
\end{tabular}
\end{center}
\caption{ The Borel parameters, continuum threshold parameters, energy scales of the QCD spectral densities,  pole contributions, and the contributions of the vacuum condensates of dimension $10$  for the ground state hidden-charm tetraquark states. }\label{BorelP}
\end{table}

\begin{table}
\begin{center}
\begin{tabular}{|c|c|c|c|c|c|c|c|c|}\hline\hline
$Z_c$($X_c$)                                                            & $J^{PC}$  & $M_Z (\rm{GeV})$   & $\lambda_Z (\rm{GeV}^5) $             \\ \hline

$[uc]_{S}[\overline{dc}]_{S}$                                           & $0^{++}$  & $3.88\pm0.09$      & $(2.07\pm0.35)\times 10^{-2}$           \\

$[uc]_{A}[\overline{dc}]_{A}$                                           & $0^{++}$  & $3.95\pm0.09$      & $(4.49\pm0.77)\times 10^{-2}$           \\

$[uc]_{\tilde{A}}[\overline{dc}]_{\tilde{A}}$                           & $0^{++}$  & $3.98\pm0.08$      & $(4.30\pm0.63)\times 10^{-2}$           \\

$[uc]_{V}[\overline{dc}]_{V}$                                           & $0^{++}$  & $4.65\pm0.09$      & $(1.35\pm0.22)\times 10^{-1}$           \\

$[uc]_{\tilde{V}}[\overline{dc}]_{\tilde{V}}$                           & $0^{++}$  & $5.35\pm0.09$      & $(4.87\pm0.51)\times 10^{-1}$             \\

$[uc]_{P}[\overline{dc}]_{P}$                                           & $0^{++}$  & $5.49\pm0.09$      & $(2.11\pm0.21)\times 10^{-1}$       \\ \hline

$[uc]_S[\overline{dc}]_{A}-[uc]_{A}[\overline{dc}]_S$                   & $1^{+-}$  & $3.90\pm0.08$      & $(2.09\pm0.33)\times 10^{-2}$        \\

$[uc]_{A}[\overline{dc}]_{A}$                                           & $1^{+-}$  & $4.02\pm0.09$      & $(3.00\pm0.45)\times 10^{-2}$           \\

$[uc]_S[\overline{dc}]_{\widetilde{A}}-[uc]_{\widetilde{A}}[\overline{dc}]_S$     & $1^{+-}$   & $4.01\pm0.09$    & $(3.02\pm0.45)\times 10^{-2}$    \\

$[uc]_{\widetilde{A}}[\overline{dc}]_{A}-[uc]_{A}[\overline{dc}]_{\widetilde{A}}$ & $1^{+-}$   & $4.02\pm0.09$    & $(6.09\pm0.90)\times 10^{-2}$    \\

$[uc]_{\widetilde{V}}[\overline{dc}]_{V}+[uc]_{V}[\overline{dc}]_{\widetilde{V}}$ & $1^{+-}$   & $4.66\pm0.10$    & $(1.18\pm0.21)\times 10^{-1}$    \\

$[uc]_{V}[\overline{dc}]_{V}$                                           & $1^{+-}$  & $5.46\pm0.09$      & $(1.72\pm0.17)\times 10^{-1}$           \\

$[uc]_P[\overline{dc}]_{V}+[uc]_{V}[\overline{dc}]_P$                   & $1^{+-}$  & $5.45\pm0.09$      & $(1.87\pm0.19)\times 10^{-1}$        \\
\hline

$[uc]_S[\overline{dc}]_{A}+[uc]_{A}[\overline{dc}]_S$                   & $1^{++}$  & $3.91\pm0.08$      & $(2.10\pm0.34)\times 10^{-2}$        \\

$[uc]_S[\overline{dc}]_{\widetilde{A}}+[uc]_{\widetilde{A}}[\overline{dc}]_S$     & $1^{++}$   & $4.02\pm0.09$    & $(3.01\pm0.45)\times 10^{-2}$    \\

$[uc]_{\widetilde{V}}[\overline{dc}]_{V}-[uc]_{V}[\overline{dc}]_{\widetilde{V}}$ & $1^{++}$   & $4.08\pm0.09$    & $(3.67\pm0.67)\times 10^{-2}$    \\

$[uc]_{\widetilde{A}}[\overline{dc}]_{A}+[uc]_{A}[\overline{dc}]_{\widetilde{A}}$ & $1^{++}$   & $5.19\pm0.09$    & $(2.12\pm0.24)\times 10^{-1}$    \\

$[uc]_P[\overline{dc}]_{V}-[uc]_{V}[\overline{dc}]_P$                   & $1^{++}$  & $5.46\pm0.09$      & $(1.89\pm0.19)\times 10^{-1}$       \\
\hline

$[uc]_{A}[\overline{dc}]_{A}$                                           & $2^{++}$  & $4.08\pm0.09$      & $(4.67\pm0.68)\times 10^{-2}$      \\

$[uc]_{V}[\overline{dc}]_{V}$                                           & $2^{++}$  & $5.40\pm0.09$      & $(2.32\pm0.25)\times 10^{-1}$      \\
\hline\hline
\end{tabular}
\end{center}
\caption{ The masses and pole residues of the ground state hidden-charm tetraquark states. }\label{mass-Table}
\end{table}

\begin{table}
\begin{center}
\begin{tabular}{|c|c|c|c|c|c|c|c|c|}\hline\hline
$Z_c$($X_c$)                                                            & $J^{PC}$  & $M_Z (\rm{GeV})$   & Assignments        &$Z_c^\prime$ ($X_c^\prime$)      \\ \hline

$[uc]_{S}[\overline{dc}]_{S}$                                           & $0^{++}$  & $3.88\pm0.09$      & ?\,$X(3860)$       &       \\

$[uc]_{A}[\overline{dc}]_{A}$                                           & $0^{++}$  & $3.95\pm0.09$      & ?\,$X(3915)$       & \\

$[uc]_{\tilde{A}}[\overline{dc}]_{\tilde{A}}$                           & $0^{++}$  & $3.98\pm0.08$      &                    & \\

$[uc]_{V}[\overline{dc}]_{V}$                                           & $0^{++}$  & $4.65\pm0.09$      &                    & \\

$[uc]_{\tilde{V}}[\overline{dc}]_{\tilde{V}}$                           & $0^{++}$  & $5.35\pm0.09$      &                    &  \\

$[uc]_{P}[\overline{dc}]_{P}$                                           & $0^{++}$  & $5.49\pm0.09$      &                    &  \\ \hline

$[uc]_S[\overline{dc}]_{A}-[uc]_{A}[\overline{dc}]_S$                   & $1^{+-}$  & $3.90\pm0.08$      & ?\,$Z_c(3900)$      &?\,$Z_c(4430)$    \\

$[uc]_{A}[\overline{dc}]_{A}$                                           & $1^{+-}$  & $4.02\pm0.09$      & ?\,$Z_c(4020/4055)$ &?\,$Z_c(4600)$        \\

$[uc]_S[\overline{dc}]_{\widetilde{A}}-[uc]_{\widetilde{A}}[\overline{dc}]_S$     & $1^{+-}$   & $4.01\pm0.09$    & ?\,$Z_c(4020/4055)$ &?\,$Z_c(4600)$     \\

$[uc]_{\widetilde{A}}[\overline{dc}]_{A}-[uc]_{A}[\overline{dc}]_{\widetilde{A}}$ & $1^{+-}$   & $4.02\pm0.09$    & ?\,$Z_c(4020/4055)$ &?\,$Z_c(4600)$    \\

$[uc]_{\widetilde{V}}[\overline{dc}]_{V}+[uc]_{V}[\overline{dc}]_{\widetilde{V}}$ & $1^{+-}$   & $4.66\pm0.10$    & ?\,$Z_c(4600)$      &    \\

$[uc]_{V}[\overline{dc}]_{V}$                                           & $1^{+-}$  & $5.46\pm0.09$      &                    &  \\

$[uc]_P[\overline{dc}]_{V}+[uc]_{V}[\overline{dc}]_P$                   & $1^{+-}$  & $5.45\pm0.09$      &                    &  \\
\hline

$[uc]_S[\overline{dc}]_{A}+[uc]_{A}[\overline{dc}]_S$                   & $1^{++}$  & $3.91\pm0.08$      & ?\,$X(3872)$       &   \\

$[uc]_S[\overline{dc}]_{\widetilde{A}}+[uc]_{\widetilde{A}}[\overline{dc}]_S$     & $1^{++}$   & $4.02\pm0.09$    &?\,$Z_c(4050)$ &   \\

$[uc]_{\widetilde{V}}[\overline{dc}]_{V}-[uc]_{V}[\overline{dc}]_{\widetilde{V}}$ & $1^{++}$   & $4.08\pm0.09$    &?\,$Z_c(4050)$ &    \\

$[uc]_{\widetilde{A}}[\overline{dc}]_{A}+[uc]_{A}[\overline{dc}]_{\widetilde{A}}$ & $1^{++}$   & $5.19\pm0.09$    &               & \\

$[uc]_P[\overline{dc}]_{V}-[uc]_{V}[\overline{dc}]_P$                   & $1^{++}$  & $5.46\pm0.09$      &                    &  \\
\hline

$[uc]_{A}[\overline{dc}]_{A}$                                           & $2^{++}$  & $4.08\pm0.09$      &?\,$Z_c(4050)$      & \\

$[uc]_{V}[\overline{dc}]_{V}$                                           & $2^{++}$  & $5.40\pm0.09$      &                    & \\
\hline\hline
\end{tabular}
\end{center}
\caption{ The possible assignments of the ground state hidden-charm tetraquark states, the isospin limit is implied. }\label{Assignments-Table}
\end{table}

\section{Conclusion}
In this article,  we take the pseudoscalar, scalar, axialvector, vector, tensor charmed (anti)diquark operators as the basic constituents, and construct
  the scalar, axialvector and tensor hidden-charm tetraquark currents to study the  mass spectrum of the ground state hidden-charm tetraquark states  with
the QCD sum rules in a comprehensive way. In calculations,   we carry out the operator product expansion up to the vacuum condensates of dimension $10$ to obtain the QCD spectral densities, and  use the energy scale formula $\mu=\sqrt{M^2_{X/Y/Z}-(2{\mathbb{M}}_c)^2}$ to determine the ideal energy scales.
The present predictions support
assigning the $X(3860)$ to be the $[qc]_{S}[\overline{qc}]_{S}$ type scalar tetraquark state with the $J^{PC}=0^{++}$,
assigning the $X(3915)$  to be the  $[qc]_{A}[\overline{qc}]_{A}$   type scalar tetraquark state with the $J^{PC}=0^{++}$,
assigning the $X(3872)$ to be the $[qc]_S[\overline{qc}]_{A}+[qc]_{A}[\overline{qc}]_S$  type axialvector tetraquark state with the $J^{PC}=1^{++}$,
assigning the $Z_c(3900)$ to be the $[uc]_S[\overline{dc}]_{A}-[uc]_{A}[\overline{dc}]_S$ type axialvector tetraquark state with the $J^{PC}=1^{+-}$,
assigning the $Z_c(4020)$ and $Z_c(4055)$ to be the $[uc]_{A}[\overline{dc}]_{A}$ type,
$[uc]_S[\overline{dc}]_{\widetilde{A}}-[uc]_{\widetilde{A}}[\overline{dc}]_S$ type or
$[uc]_{\widetilde{A}}[\overline{dc}]_{A}-[uc]_{A}[\overline{dc}]_{\widetilde{A}}$ type axialvector tetraquark states with the $J^{PC}=1^{+-}$,
assigning the $Z_c(4600)$ to be the $[uc]_{\widetilde{V}}[\overline{dc}]_{V}+[uc]_{V}[\overline{dc}]_{\widetilde{V}}$ type axialvector tetraquark state with the $J^{PC}=1^{+-}$, or the first radial excited state of the $Z_c(4020)$,
assigning the $Z_c(4050)$ to be the $[uc]_S[\overline{dc}]_{\widetilde{A}}+[uc]_{\widetilde{A}}[\overline{dc}]_S$ type or
$[uc]_{\widetilde{V}}[\overline{dc}]_{V}-[uc]_{V}[\overline{dc}]_{\widetilde{V}}$ type axialvector tetraquark state with the $J^{PC}= 1^{++}$  or
$[uc]_{A}[\overline{dc}]_{A}$ type tensor tetraquark state with  the $J^{PC}=2^{++}$,
assigning the $Z_c(4430)$ to be  the first radial excited state of the $Z_c(3900)$.
More experimental data and theoretical work are still needed to make  unambiguous assignments.
There are no rooms to accommodate the $X(3940)$, $X(4160)$, $Z_c(4100)$, $Z_c(4200)$ in the scenario of tetraquark  states without fine tuning. The $X(3940)$ and $X(4160)$ may be  the conventional $\eta_c(\rm3S)$ and $\eta_c(\rm 4S)$ states  with the $J^{PC}=0^{-+}$, respectively. While the $Z_c(4100)$ may be a mixing scalar  tetraquark state with the $J^{PC}=0^{++}$, the $Z_c(4200)$ may be  an axialvector color-octet-color-octet  type tetraquark state with the $J^{PC}=1^{+-}$.
The predicted tetraquark masses can be confronted to the experimental data in the future at the BESIII, LHCb, Belle II,  CEPC, FCC, ILC.

\section*{Acknowledgements}
This  work is supported by National Natural Science Foundation, Grant Number  11775079.

\end{document}